\documentclass[a4paper,fleqn]{cas-dc}

\usepackage[numbers,sort&compress]{natbib}
\usepackage{url}
\usepackage{graphicx}

\usepackage{caption}
\usepackage{subcaption}
\usepackage[utf8]{inputenc}
\usepackage{amssymb}
\usepackage{amsmath}
\usepackage{dcolumn}
\usepackage{bm}
\usepackage{makecell}
\usepackage[T1]{fontenc}
\usepackage{etoolbox}
\usepackage{xcolor}
\usepackage{float}
\usepackage{hyperref}
\usepackage{cleveref}
\usepackage{comment}
\def\tsc#1{\csdef{#1}{\textsc{\lowercase{#1}}\xspace}}
\tsc{WGM}
\tsc{QE}
\begin{document}
\let\WriteBookmarks\relax
\def\floatpagepagefraction{1}
\def\textpagefraction{.001}

%\shorttitle{Photonic-glass coatings for colored BIPV}    

\shortauthors{S. Genc et al.}

\title[mode = title]{Engineering Color and Angular Stability in Building-Integrated Photovoltaics Using Metal-Oxide Photonic Glasses}
%\title[mode = title]{Colored building-integrated photovoltaics with metal-oxide photonic glasses: reachable colors and the limits of angular stability}

\author[1,2]{Sinan Genc}%[<options>]

\credit{Software, Validation, Formal analysis,  Data curation, Writing - original draft, Writing - review \& editing, Visualization}

\affiliation[1]{organization={Abdullah Gül University},
            addressline={Department of Electrical-Electronics Engineering}, 
            city={Kayseri},
%          citysep={}, % Uncomment if no comma needed between city and postcode
            postcode={38080}, 
            country={Turkey}}
\affiliation[2]{organization={Middle East Technical University},
            addressline={Department of Physics}, 
            city={Ankara},
%          citysep={}, % Uncomment if no comma needed between city and postcode
            postcode={06800}, 
            country={Turkey}}
            
\author[3]{Talat Ozden}%[]

\credit{Conceptualization, Resources, Writing - review \& editing, Supervision, Project administration, Funding acquisition}

\affiliation[3]{organization={ODTÜ-GÜNAM},
            addressline={Center for Solar Energy Research and Applications}, 
            city={Ankara},
%          citysep={}, % Uncomment if no comma needed between city and postcode
            postcode={06800}, 
            country={Turkey}}

\affiliation[4]{organization={Middle East Technical University},
            addressline={Department of Micro and Nanotechnology}, 
            city={Ankara},
%          citysep={}, % Uncomment if no comma needed between city and postcode
            postcode={06800}, 
            country={Turkey}}

\cortext[1]{Corresponding author}

\author[2,3,4]{Alpan Bek}%[<options>]

\cormark[1]

\ead{bek@metu.edu.tr}

\credit{Conceptualization, Methodology, Validation, Formal analysis, Investigation, Data curation, Writing - original draft, Writing - review \& editing, Supervision}

\begin{abstract}
Coloring building-integrated photovoltaics poses a dilemma: absorptive pigments cost photocurrent, while ordered photonic structures are iridescent. Amorphous photonic glasses, disordered packings of high-index nanospheres, are proposed to avoid both, yet their angle-independent color has rarely been quantified on a realistic module stack. A non-absorbing scatterer needs a high index with visible transparency, which the Moss rule caps near $2.35$. Coupling Mie scattering, the Percus--Yevick structure factor, the exact phase function and two-flux transfer with absorption and refraction at the front glass, we fix four target colors that a crowd-sourced naming survey calls unambiguously blue, green, gold and terracotta, and map the recipes reaching them across four metal oxides, four encapsulants, two cell technologies and wide ranges of particle size, packing fraction and layer number. All four are reached in a laminated three-layer coating, to within $\Delta E_{00}$ of 0.8 to 2.5, but not from one oxide: transparent TiO$_2$ renders the blue and only pale warm tones, the green, gold and terracotta require the blue-band absorption of V$_2$O$_5$, and the low-contrast oxides turn grey at the warm end. The encapsulant governs the color and the cell technology does not. Disorder does not confer angle independence: the colors hold for an observer moving under a fixed sun but drift by $\Delta E_{00}$ of 5 to 18 over the day, terracotta least, and a façade lit from above and viewed from below shifts every color towards the blue. Oxide, layer number and façade geometry must therefore be designed together.
\end{abstract}

%\centering
%\includegraphics[width=13cm,keepaspectratio]{GA.pdf}

%\item Four crowd-named façade colors are reachable in a laminated photonic glass.
%\item No single oxide reaches all four; warm tones need blue-band absorption.
%\item Refraction at the front glass compresses the angles that set the color.
%\item Seen from below, a façade lit from above shifts every color towards blue.
%\item A recipe designed before lamination is a different color once encapsulated.

%\nocite{*}

% Keywords

\maketitle

\section{Introduction}\label{sec:intro}

Building-integrated photovoltaics (BIPV) are widely regarded as an
indispensable technology for decarbonizing the built environment and
realizing zero-energy buildings, because they convert otherwise passive
building surfaces into generating assets without competing for additional
land \citep{Block2024ColouringReview, Mao2025StructuralColorReview}. Yet the
deployed capacity of BIPV remains far below its technical potential, and the
dominant barrier is aesthetic rather than energetic: conventional silicon
modules present a dark blue-to-black appearance, dictated by the requirement
of maximal broadband absorption, that architects, planning authorities, and
end users frequently find unacceptable on visible façades and
heritage-sensitive sites \citep{Pearce2024EfficiencyLimits,
Block2024ColouringReview}. This is not a matter of taste alone. Urban
planning authorities in several countries already restrict the use of black
photovoltaic surfaces on building façades on visual-impact grounds, while
survey evidence shows that offering color markedly improves public
acceptance, with roughly 81\% of respondents expressing a supportive attitude
toward colored BIPV \citep{ColoredBIPVReview2026}. When color and seamless
integration are offered, acceptance rises sharply, and occupants are
frequently willing to pay a premium for modules that enhance the visual
appeal of their building \citep{Block2024ColouringReview,
ConsumerFacadePreference2024}. Color is therefore not a cosmetic add-on but a
functional prerequisite for adoption, and, because façade area vastly exceeds
available roof area in dense urban settings, removing this aesthetic barrier
directly unlocks a large fraction of the built environment for solar
harvesting \citep{Pearce2024EfficiencyLimits,Li2025Nexus}.

The colors that a façade should present are, moreover, not arbitrary. A
substantial body of work in architecture and environmental psychology has
quantified exterior color preference and its dependence on cultural context,
building typology, and observer demographics \citep{ColorHarmonyFacade,
ConsumerFacadePreference2024}. Field surveys consistently identify muted,
low-chroma tones, rather than vivid saturated hues, as the palette most
compatible with the built environment and most frequently specified by
architects and residents, with terracotta and brick reds, greens, warm ochres
and golds, and desaturated blues and greys recurring across studies
\citep{ConsumerFacadePreference2024}. The same small set reappears in the
BIPV context specifically, where green and terracotta are the archetypal
target colors of both market products and field-tested façades
\citep{ThreeColorFacade2026}. Any coloring technology intended for real
deployment must therefore be assessed not against its ability to produce
arbitrary laboratory colors, but against its ability to reach this
architecturally relevant target set. We adopt this principle as the
organizing device of the present study: rather than reporting whatever colors
the model happens to produce, we fix a small number of façade-relevant target
colors at the outset and ask, systematically, which physical recipes reach
them and under what conditions those colors hold.

Existing PV coloring strategies fall into two families, each with an
intrinsic limitation. The first, absorptive coloring, applies pigments or
dyes that produce color by subtracting part of the visible spectrum; the
absorbed photons are lost to photocurrent, so saturation and efficiency are
in direct tension. Peharz and Ulm~\cite{PeharzUlm2018} showed through fundamental optical
calculations that, while highly saturated monochromatic colors can in
principle be realized at power losses below about seven percent, standardized
façade colors incur substantially higher losses, with warm hues such as
yellow, orange, and red penalized roughly twice as heavily as blue, green,
and dark tones. The second family, structural coloring, produces color
through wavelength-selective reflection from ordered photonic architectures
such as dielectric Bragg stacks and Fabry--P\'erot filters, and largely
avoids parasitic absorption \citep{Zhang2015Perovskite, Xu2024Multilayer}.
However, coloring based on long-range periodic order is inherently
iridescent: the Bragg condition ties the reflected wavelength to the angle of
incidence and observation, so the perceived color shifts as the sun traverses
the sky and as the viewer moves. For a building façade, which is seen from
many directions and illuminated at continuously varying angles throughout the
day, this angular color instability is architecturally undesirable, and it is
precisely the property that architects seeking a stable, uniform surface wish
to avoid \citep{Xu2022AngleInsensitive}.

Amorphous photonic materials, or ``photonic glasses,'' are widely proposed as
the route that retains the low-absorption advantage of structural color while
removing its angular dependence. In a disordered but short-range-correlated
packing of dielectric spheres there is no long-range periodicity and hence no
Bragg diffraction; color instead arises from the single-particle Mie
resonance together with an isotropic contribution from the nearest-neighbor
correlation \citep{Schertel2019StructuralColors, Hwang2020AngleIndependent}.
Because the structure factor depends only on the magnitude of the momentum
transfer, the correlated enhancement itself carries no angular preference,
and from this it is commonly inferred that the resulting color is
non-iridescent. The feasibility of the approach for photovoltaics has been
established: Li et al.~\cite{Li2022PhotonicGlass} showed both theoretically and
experimentally that a sprayed layer of monodisperse dielectric nanospheres
can color silicon cells with negligible parasitic absorption, obtaining blue,
green, and purple cells at a power conversion efficiency of 21.5\% against
22.6\% for the black reference, and subsequent theoretical work has mapped
the efficiency and lightness trade-off achievable with non-absorbing
high-index spheres \citep{Li2023PhotonicGlassCovers, LiMa2022EfficiencyLimit}. The same group has since carried the concept into an industrial module
process \citep{Li2025Nexus}: silica spheres are packed, infiltrated with an
acrylate and UV cured, the solid is ground into pigment grains of tens of
micrometres, and the grains, dispersed in a UV-curable resin, are printed on
the inner face of the cover glass and laminated with EVA against
interdigitated back-contact cells. The resulting light blue, light cyan and
greyish-white modules reach power conversion efficiencies of 21.21, 20.25
and 19.99\% against 22.40\% for the black reference.

A second and quite different route through the same design space has recently been demonstrated by Song et al.~\cite{Song2026SubMonolayer}, and the contrast with it locates the present work precisely. They slot-die coat a sub-monolayer of silicon nanospheres, 110 to 200\,nm in diameter, onto the inner face of the module cover glass, laminate it against the cell with EVA, and report blue, green and yellowish mini-modules at relative photocurrent losses of about 10 to 20\%. The particles are encapsulated in a polymer shell so that neighbors remain optically independent, since clustering broadens the resonance and desaturates the color. Their coloration therefore rests on the resonance of the individual scatterer, and the short-range positional order that defines a photonic glass is suppressed rather than exploited; the correlated packing considered here is the opposite limit of the same problem. The two regimes differ in what they can offer. A dilute array can use a material of very high index, silicon at $n \approx 3.5$, because a photon interacts with one particle and leaves, so the modest absorption of silicon in the visible is tolerable and vivid colors follow from the large contrast alone; a dense packing cannot, since a photon crosses many layers and any absorption compounds along the path, which is why the index ceiling discussed below binds here and not there. In exchange, the dense packing adds the structure factor as a second spectral filter, reaches lighter tones, with the reference recipes of this work spanning $L^*$ from 45 to 75 against the 40 to 50 reported for the sub-monolayer, can be built largely from oxides that are transparent across the visible, and offers the packing fraction and the layer number as design variables that a sub-monolayer does not possess.

 The pigment route of \cite{Li2025Nexus} occupies the opposite extreme of
index contrast. Silica and the acrylates that fill its interstices, whose
indices span 1.47 to 1.58, differ in index only slightly, so each sphere is a
weak scatterer and the color rests almost entirely on the short-range order,
whose correlation length the authors estimate from small-angle X-ray
scattering at about two sphere diameters; the weak contrast is then
compensated by thickness, with films of about 300\,$\mu$m. This choice buys
two things a high-index packing does not have. Both constituents are
transparent, so the coating costs photocurrent only through what it reflects,
and the matrix that sets the contrast is locked inside each grain, so the
color is fixed when the pigment is made rather than when the module is
laminated. The same grains appear white as a dry powder and colored once
wetted, a qualitative image of the medium dependence quantified below. What
the low-contrast route gives up is reach. Every color it reports lies at
negative $b^*$, on the blue side of neutral, and raising the matrix index
brightens the color while desaturating it, so the golds, ochres and
terracottas of the façade palette are absent. The packing considered here
sits at the high-contrast end: transparent oxides at the Moss ceiling,
embedded directly in the encapsulant, color a module in a coating of three
monolayers, below a micrometre thick, and, as shown in
Section~\ref{sec:materialswap}, the blue-absorbing V$_2$O$_5$ extends that
reach to gold and terracotta, at a photocurrent cost that non-absorbing
pigments do not incur.

What these demonstrations leave open is design and quantification. The colors reported are those the particles happen to produce rather than an architecturally specified palette, and in all three routes they omit the warm tones; the angular stability is documented by photographs at oblique tilts rather than by measured color differences; and lamination is either reported not to move the color, for the dilute array, or kept from moving it by locking the matrix inside each pigment grain, whereas for correlated oxide packings embedded directly in the encapsulant the color shifts by up to $\Delta E_{00} = 45.1$, as shown in Section~\ref{sec:materialswap}. Four questions decisive for façade deployment therefore remain open, and they organize what follows.

The first is where such a coating would actually be applied. A colored module
is only adopted if some manufacturer can produce it, and the two candidate
points of insertion are not equally plausible. Cell and module production
lines are highly optimized, tightly toleranced, and capital-intensive, and
appending a further wet-deposition step to the end of one, with the yield,
throughput and contamination risk that entails, is a difficult proposition to
put to a cell manufacturer whose margins depend on that optimization. Flat
glass production is the more tractable point of entry. The equipment required
is modest, an ultrasonic-spray nozzle at the exit of the line, the substrate
is already flat and moving, and the output is pre-colored, lamination-ready
glass that reaches the panel maker as a drop-in replacement for clear front
glass, with no added assembly step on the module line at all. The incentive
is aligned on that side as well: glass producers already supply colored and
coated architectural glazing, whereas the cell manufacturer gains little but
risk. A coloring route that can be deposited from suspension in a single
pass, as spray-deposited photonic glasses can \cite{Li2022PhotonicGlass}, is
therefore attractive for a reason that is industrial rather than optical. The same reasoning has led Song et al.~\cite{Song2026SubMonolayer} to place their coating on the inner face of the cover glass, where the glass itself serves as the protective barrier and the cell process is left untouched, and to deposit it by slot-die coating, a route equally compatible with a moving glass line; this is independent support for the choice made here.
This framing carries a direct physical consequence to which we return below:
deposition on the glass side means the coating is encapsulated before it is
ever seen, so its color must be designed in the laminated state rather than
as deposited.

The second concerns the target colors themselves. Existing photonic-glass
demonstrations report the colors their particles happen to produce, typically
blues, greens and purples, and more recently greys and whites
\citep{Li2025Nexus}, whereas the palette that matters is the one
specified by architects and supplied by the architectural glazing industry,
which is largely warm and consistently muted. Which of those tones are
reachable, and with which recipe, has not been established. The
inaccessibility of saturated red in purely dielectric disordered packings is
a known constraint of the underlying physics
\citep{Magkiriadou2014RedAbsence, Jacucci2020Limitations}, but a façade does
not call for a saturated red; it calls for muted terracottas and ochres, and
whether those fall inside the reachable region is a distinct question.

The third concerns the choice of scattering material. TiO$_2$ is the natural
first choice for the index contrast on which the Mie resonance depends, but
it is not always the preferred one: cost, photocatalytic activity at the
encapsulant interface, and the difficulty of growing large monodisperse
particles all motivate alternatives. Whether a lower-index oxide can
reproduce a given target color, at what particle size, and at what cost in
chroma, determines how much freedom a manufacturer actually has. The question
cannot be settled by comparing materials at a fixed diameter, since each
material reaches a given color on its own size scale, and to our knowledge it
has not been posed quantitatively for correlated disordered packings on a
photovoltaic stack.

The fourth, and the one we regard as most consequential, concerns the angular
claim itself. The argument as usually stated establishes only that the
correlated enhancement is isotropic in momentum transfer. It does not follow
that the perceived color is invariant, because the scattering angle
accessible to an observer changes with geometry, the single-particle phase
function retains sharp resonant structure, and the substrate and multiple
scattering both contribute. A façade is moreover illuminated by a direct beam
whose elevation changes throughout the day, superposed on diffuse skylight
whose relative weight itself grows as the sun descends, and it is viewed from
a continuum of azimuths rather than only in the plane of incidence. How far
the color holds under these conditions, and how that depends on the recipe,
has not been quantified for a photovoltaic photonic glass.

Before turning to the model, the restriction to metal oxides warrants a word,
since it might appear arbitrary. It is not. The scatterer must combine a high
refractive index, which sets the contrast that drives the Mie resonance, with
negligible absorption across the visible, without which the coating would
compete with the cell for photons. These two requirements are in tension. The
empirical Moss rule relates the sub-bandgap refractive index to the band gap
as $E_g n_0^4 \approx 95$\,eV \citep{Moss1985, Doiron2022SuperMossian}, so
that transparency throughout the visible, which demands an absorption edge
above roughly $3.1$\,eV, caps the index near $n \approx 2.35$. Materials that
surpass this bound exist, and TiO$_2$ is itself mildly super-Mossian, but the
high-index platforms that substantially exceed it, such as silicon, gallium
phosphide, and the transition-metal dichalcogenides, do so at the price of
absorption within the visible band \citep{Khurgin2022PhotonicPalette}.
Nitrides illustrate the same constraint from both sides. Titanium nitride is
not a dielectric at all: its free-carrier density of order
$10^{22}$\,cm$^{-3}$ renders it metallic in the visible, which is the origin
of its familiar golden appearance and of losses that would directly reduce
photocurrent \citep{Naik2012TiN, Naik2011OxidesNitrides}. The nitrides that
are transparent, such as AlN with a band gap of $6.2$\,eV, pay for that
transparency with an index near $2$, below every oxide considered here. Metal
oxides are the material class that sits at this ceiling while also offering
the wet-chemical routes to monodisperse nanospheres and the long-term
outdoor stability that a façade coating requires, and it is for these reasons
that the present survey is built on them. Wide-gap sulfides occupy a
comparable index range, and ZnS in particular has already been demonstrated
as a photonic-glass colorant on silicon cells \citep{Li2022PhotonicGlass};
the same framework applies to them without modification, but their
processing, encapsulant compatibility, and outdoor durability differ
sufficiently from those of the oxides that a like-for-like comparison lies
beyond the scope of this work. The contrast limitation that runs through the
results below is accordingly a property of the ceiling itself, not of the
particular four oxides chosen. Within the oxides, the one way past the
ceiling is to accept absorption, as V$_2$O$_5$ does in the blue; whether that
helps or harms the palette is one of the questions answered below.

In this work we address the questions above with a numerical model of the
structure sketched in Fig.~\ref{fig:schematic}(a), coupling Mie scattering,
the Percus--Yevick hard-sphere structure factor, the exact Mie phase
function, and multilayer radiative transfer over a photovoltaic substrate,
with the resulting color computed from the CIE~1931 standard observer under
AM1.5G illumination \citep{CIE15, ASTMG173}, chromaticity reported in the
CIE~1976 $u'v'$ diagram, and both chroma and color differences evaluated in
CIELAB with the CIEDE2000 metric \citep{LuoCuiRigg2001,
Sharma2005CIEDE2000}. The first of the four questions is settled by the
argument already given: the coating belongs on the glass side of the supply
chain, and everything that follows is computed for a coating in that state,
embedded in the encapsulant rather than exposed. The remaining three are
answered by computation. Taking four target colors, chosen from the
reachable palette by an external naming criterion, as fixed targets, we sweep
particle size, material, packing fraction, layer number, embedding medium,
substrate, and both illumination and observation geometry, and we evaluate
the resulting colors under directional, clear-sky and fully diffuse
illumination. Four results organize the roadmap that emerges. All four
targets are reached, but not from one oxide: a transparent high-index oxide
gives the cool tones and a blue-absorbing one the warm tones and green, so at
the index ceiling absorption is what extends a scattering coating to the warm
half of the palette. The surrounding medium governs the result and the cell
technology beneath does not, so a recipe developed before lamination is a
different color once encapsulated, and design belongs on the finished stack,
which is exactly the state in which glass-side deposition delivers it.
Angular behavior is a design variable rather than a property conferred by
disorder: the colors hold for an observer moving under a fixed sun but drift
as the sun moves, and they shift towards the blue in the geometry of a façade
lit from above and viewed from below. And the layer number, set here by where
all four targets keep their names, is a weak lever on stability. Together
these convert the photonic-glass concept into a design roadmap that states
not only what is achievable, but under which conditions it holds and at
which point in the supply chain it can be produced.

\section{Theory and Model}\label{sec:model}

\begin{figure*}[t]
  \centering
  \includegraphics[width=0.9\textwidth]{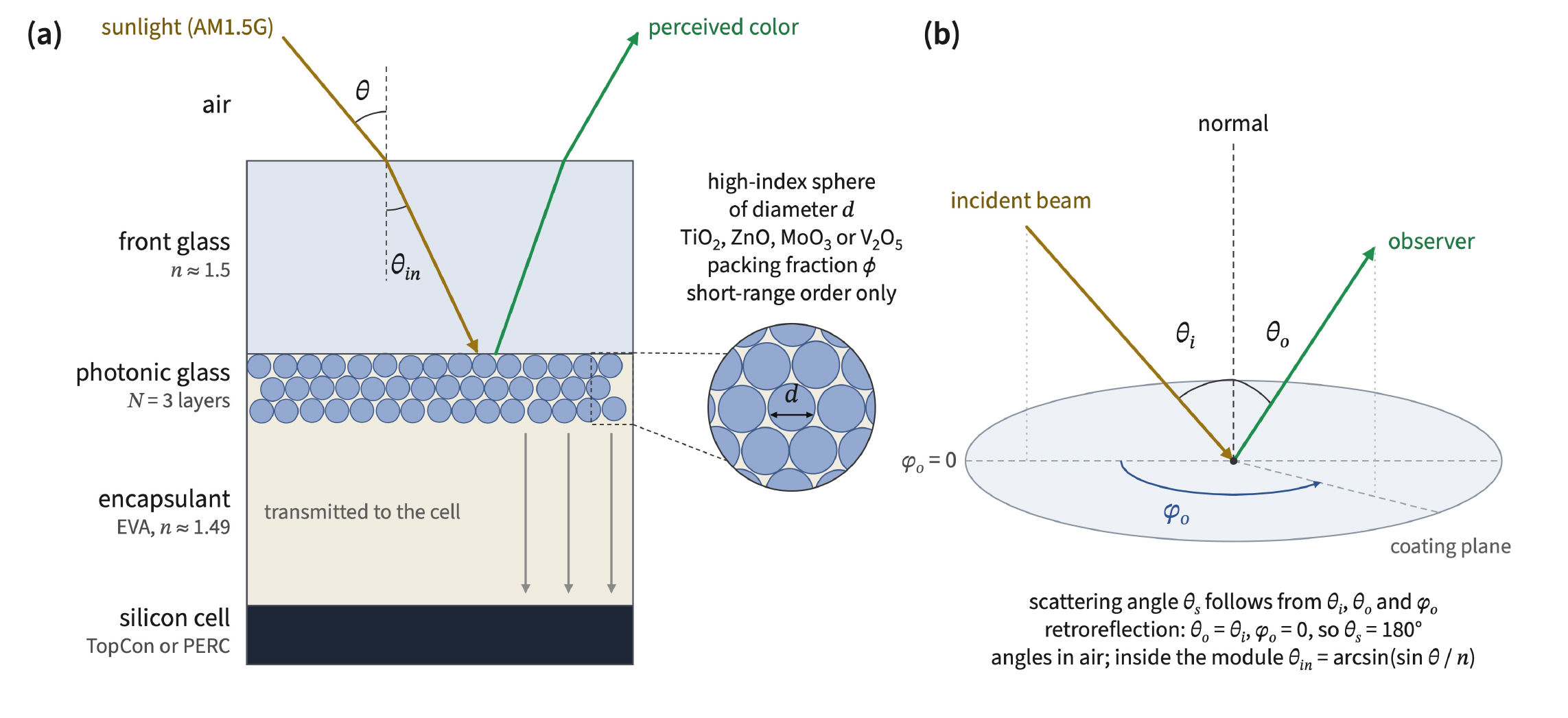}
  \caption{(a)~Structure considered in this work: a disordered, correlated
  packing of monodisperse high-index nanospheres embedded in the encapsulant
  between the front glass and the photovoltaic cell. (b)~Illumination and
  observation geometry. The polar angle $\theta$ is measured from the surface
  normal for both the incident beam and the observer, and the azimuthal angle
  $\varphi_o$ is measured in the plane of the module from the plane of incidence,
  so that $\varphi_o = 0$ places the observer on the side of the source, where
  the retroreflection direction lies. All polar angles are measured in air,
  outside the module.}
  \label{fig:schematic}
\end{figure*}

The optical response of a photonic-glass coating on a photovoltaic module is
computed here in five coupled stages: the scattering of a single high-index
nano-/micro-sphere embedded in a dielectric medium, obtained from Mie theory; the
modification of the scattered intensity by short-range positional
correlations between neighboring spheres, described by the Percus--Yevick
structure factor; the angular redistribution of the scattered light,
evaluated from the exact Mie phase function so that the fraction returned
towards the observer can be extracted for an arbitrary illumination and
observation geometry; the propagation of the resulting reflectance through a
multilayer stack terminating on the photovoltaic substrate; and the weighting
of these directional responses by a model of the incident sky, followed by
conversion of the spectral reflectance to a perceived color. The structure to
which this applies is sketched in Fig.~\ref{fig:schematic}(a), and the
angular conventions used throughout are defined in
Fig.~\ref{fig:schematic}(b). This section sets out the physical content of
each stage and the few relations needed to follow the results; the complete
set of equations, the numerical conventions, and the verification of the
implementation against independent references are collected in the
Supplementary Material.

\subsection{Scatterer, medium and substrate}\label{sec:materials}

The elementary scatterer is a homogeneous sphere of radius $r$ and complex
refractive index $m = n + ik$, immersed in a non-absorbing medium of index
$n_{\mathrm{med}}$. Its interaction with light of vacuum wavelength $\lambda$
is governed by the size parameter
\begin{equation}
  x = \frac{2\pi\, n_{\mathrm{med}}\, r}{\lambda},
  \label{eq:sizeparam}
\end{equation}
and by the relative refractive index
\begin{equation}
  \tilde m = \frac{m}{n_{\mathrm{med}}} ,
  \label{eq:relindex}
\end{equation}
which is the contrast that drives the Mie resonance. The scattering
coefficients $a_n$ and $b_n$ are obtained from the standard Mie expansion in
vector spherical harmonics \citep{BohrenHuffman1983}, from which the
scattering and extinction efficiencies $Q_{\mathrm{sca}}$ and
$Q_{\mathrm{ext}}$ follow, with the absorption efficiency as their
difference. For a sphere with $k = 0$ across the visible the absorption
efficiency vanishes and all light removed from the incident beam is
redistributed rather than dissipated, which is the condition under which the
coating colors the module without competing with the cell for photons.

The scatterers are monodisperse metal-oxide colloidal nanospheres, and four candidate oxides are considered: TiO$_2$, ZnO, MoO$_3$ and V$_2$O$_5$. Their dispersion $n(\lambda)$ and extinction $k(\lambda)$ are taken from published visible-range determinations, compiled through~\cite{Polyanskiy2024RII} where available, and interpolated onto the colorimetric wavelength grid. For TiO$_2$ we use the anatase
data of \cite{Jolivet2023TiO2}, the phase accessible to the
room-temperature wet-chemical routes relevant here, and for ZnO the
single-crystal ordinary-ray dispersion of \cite{Bond1965ZnO}, extrapolated from its lower limit of $450$\,nm to $380$\,nm, with $k = 0$ since the band gap lies below the visible. MoO$_3$ requires a
compromise: the published constants differ by about twenty percent between
the crystalline $\alpha$ phase \cite{Lajaunie2013MoO3}, with
$n \approx 2.35$ at $400$\,nm along its lowest-index axis, and sputtered films
\cite{Stelling2017}, with $n \approx 1.96$, and since the phase of a
wet-chemically grown nanosphere is not guaranteed to be either, we adopt
the arithmetic mean of the two sets, in $n$ and in $k$. V$_2$O$_5$ is absent from the database of \cite{Polyanskiy2024RII}, and we
build its constants from the sol-gel film determination of Benmoussa et
al.~\cite{Benmoussa2003V2O5}, the preparation route closest to the wet-chemical
synthesis considered here. A Kramers--Kronig consistent Tauc--Lorentz
oscillator \cite{Jellison1996TaucLorentz} is fitted jointly to the Sellmeier
dispersion reported there, which it reproduces to within $0.02$ in $n$ above
$600$\,nm, and to its Tauc absorption, with a band gap of $2.3$\,eV; the
weak sub-gap tail of about $3\times10^{3}$\,cm$^{-1}$ shown in the same work
is added to $k$. The oscillator thus supplies $n$ within the absorption band,
where it was not measured, and $k$ below $420$\,nm, where the measured
absorption ends. Sputtered films
\cite{Benmoussa1995V2O5, Kang2013V2O5}, as compiled in
\cite{Schneider2020Vanadium}, show indices about $0.15$ to $0.25$ higher in
the red, an indication of the spread between preparation routes. TiO$_2$ offers the highest index, falling from
$n \approx 2.87$ at $380$\,nm to $n \approx 2.43$ at $780$\,nm, with no measurable
extinction across the visible in that determination, so it approaches the ideal
non-absorbing colorant. Its photocatalytic activity would attack the surrounding polymer over the
lifetime of a module, but this is a solved problem in the pigment industry:
TiO$_2$ particles are routinely passivated with a silica shell of a few nanometers. Such a shell is not optically negligible, since it lowers the contrast of each sphere and shifts the rendered color, so it is best treated as a design variable absorbed by retuning the core diameter; we consider bare spheres here and leave the shell to a separate study. ZnO and MoO$_3$ provide markedly lower contrast, with
$n$ settling near $1.95$ in the red; ZnO is transparent across the
visible, whereas MoO$_3$ retains a weak violet tail, $k \approx 0.07$ at
$380$\,nm falling below $0.01$ by $500$\,nm. V$_2$O$_5$ lies
slightly below TiO$_2$ in index in the red, $n \approx 2.27$ against $2.45$ at
$700$\,nm, and above it in the blue, $n \approx 3.0$ against $2.77$ at
$400$\,nm, but absorbs strongly below $500$\,nm, with $k \approx 0.97$ at
$380$\,nm falling to $0.05$ at $500$\,nm, and retains a weak sub-gap tail,
$k$ of $0.007$ to $0.019$ across the rest of the visible, so it is not an
absorption-free colorant. It is retained here because that behavior
proves consequential for the warm and green target colors, and because the resulting trade-off between index contrast and blue-band transparency is itself one of
the design constraints this work sets out to quantify. The optical constants
of all four are plotted in Fig.~S1 of the Supplementary Material, and the
sensitivity of the results to them is quantified there as well.

The spheres are embedded either in air ($n_{\mathrm{med}} = 1$) or in a
polymer encapsulant, as sketched in Fig.~\ref{fig:schematic}(a).
Ethylene-vinyl acetate (EVA) is the reference encapsulant throughout, with a
weakly dispersive index rising from $1.487$ in the red to $1.501$ in the
violet. The choice remains representative of current module practice, in
which the single EVA layer is increasingly replaced for TopCon architectures
by the co-extruded EVA-POE-EVA structure known as EPE: the polyolefin
interlayer improves moisture ingress and resistance to potential-induced
degradation, but the layer in contact with the front glass, and therefore
with the coating, is EVA in both cases, and the refractive index of POE
itself differs from that of EVA by less than one percent. Three further
encapsulants, POE, PVB and TPU, are examined in the Supplementary Material.
Because the contrast that drives the resonance is the ratio of
Eq.~\eqref{eq:relindex} rather than the sphere index alone, embedding a
sphere in a polymer rather than in air reduces that contrast substantially: a
TiO$_2$ sphere with $n \approx 2.5$ presents a contrast of $2.5$ in air but only about $1.7$ in EVA. This single factor is responsible for much of what
follows, and it is also why the unlaminated and laminated states cannot be
treated as interchangeable.

Beneath the coating, the substrate enters through the spectral reflectance
$R_{\mathrm{sub}}(\lambda)$ of the photovoltaic cell as seen through the
encapsulant. Two silicon cell technologies are used, PERC and TopCon, each
represented by a parametric reflectance curve with the shallow minimum near
$600$\,nm imposed by its antireflection coating, rising towards both ends of
the visible, which gives the characteristic dark blue-violet appearance of
an uncolored module; both curves are plotted in Fig.~S2. Replacing them by
a physics-based alternative, a a textured silicon cell under a quarter-wave ($73$\,nm)
silicon nitride coating computed from tabulated optical constants
\cite{Green2008Si, Luke2015Si3N4}, moves the reference colors by
$\Delta E_{00}$ of $0.2$ to $3.4$ (Supplementary Material), so the
conclusions do not rest on the parametric form. A black substrate with
$R_{\mathrm{sub}} = 0.04$, combined with air as the medium, serves as an
idealized maximally absorbing reference against which the effect of a
realistic module stack can be isolated; it also represents the coating as it
would be inspected before lamination. The three combinations examined
throughout are therefore air over black, EVA over PERC, and EVA over TopCon.

\subsection{Correlated scattering and the viewing geometry}\label{sec:phase}

A photonic glass is not a dilute suspension of independent scatterers but a
dense, jammed packing in which each sphere is surrounded by a well-defined
shell of nearest neighbors at an approximately fixed separation. This
short-range positional correlation, in the absence of any long-range
periodicity, is the physical origin of the structural color: it imposes a
preferred interparticle spacing that enhances scattering at a particular
wavelength while remaining statistically isotropic. The correlation is
captured by the static structure factor $S(q)$, which multiplies the
single-particle scattered intensity. We use the hard-sphere structure factor
in the Percus--Yevick approximation, for which the direct correlation
function admits the closed analytic solution of \cite{Wertheim1963} in the
form given by \cite{AshcroftLekner1966}; the algebra is reproduced in the
Supplementary Material. In the dilute limit $\phi \to 0$ one recovers
$S(q) \to 1$, the independent-scatterer result. As the packing fraction
increases towards the dense packings characteristic of a real photonic glass,
the first maximum of $S(q)$ grows and sharpens, its position remaining of
order $qd \approx 2\pi$, that is, at a momentum transfer set by the mean
nearest-neighbor separation. It is this peak that concentrates
backscattering into a narrow spectral band and controls the chroma of the
perceived color; the curves are plotted across the range used here in
Fig.~S3. Equally important is what happens away from the peak, since at a
packing fraction near close packing the structure factor falls well below
unity over most of the backward hemisphere, so a dense correlated packing
suppresses scattering outside its resonant band as strongly as it enhances
scattering within it. Packing fractions from $\phi = 0.39$ to $\phi = 0.64$
are examined throughout this work, spanning a dilute disordered packing to the
random close-packed limit. The packing fraction enters the reflectance a
second time, through the optical thickness of a monolayer, as set out in
Section~\ref{sec:stack}.

The perceived color is set not by the total scattering efficiency but by the
fraction of light returned towards the observer, so the angular distribution
of the scattered intensity must be evaluated explicitly. We use the exact Mie
phase function $p(\theta)$ constructed from the same coefficients $a_n$ and
$b_n$, rather than the Henyey--Greenstein approximation parameterized by the
asymmetry factor $g$: at the size parameters of interest the phase function
develops sharp resonant structure that a single-parameter form cannot
reproduce, and that structure is precisely what carries the color. The same
angular signatures have been analyzed and experimentally verified for
dielectric microspheres in aqueous media in earlier work
\citep{Genc2023RSOS, Genc2024SNA}. Correlated scattering then enters through
the structure factor evaluated at the momentum transfer associated with each
scattering angle,
\begin{equation}
  q(\theta) = 2k\,\sin(\theta/2),
  \qquad k = \frac{2\pi n_{\mathrm{med}}}{\lambda},
  \label{eq:qtheta}
\end{equation}
so that the angular response of the packing is the product
$p(\theta)\,S\!\big(q(\theta)\big)$. Two evaluations of this response are
used. The first integrates it over the entire backward hemisphere and
describes the diffuse appearance of the coating without reference to a
particular observer. The second, appropriate to a specified viewing
configuration, averages it over a narrow band centered on the single
scattering angle fixed by the illumination and observation directions. Neither
carries a projection factor: colors are computed as reflectance factors,
relative to a white diffuser under the same illumination, for which the
reduced projected irradiance at oblique incidence cancels. Both are written
out in the Supplementary Material.

The scattering angle entering the directional evaluation follows from the
geometry of Fig.~\ref{fig:schematic}(b). For an incident direction at polar
angle $\theta_i$ and an observation direction at polar angle $\theta_o$ and
azimuth $\varphi_o$ measured from the plane of incidence,
\begin{equation}
  \cos\theta_s = -\big(\sin\theta_i\,\sin\theta_o\,\cos\varphi_o
                     + \cos\theta_i\,\cos\theta_o\big),
  \label{eq:thetascatter}
\end{equation}
which reduces, for an observer in the plane of incidence, to
$\theta_s = 180^{\circ} - \lvert\theta_o - \theta_i\rvert$. The
retroreflection condition $\theta_i = \theta_o$, $\varphi_o = 0$ therefore gives
$\theta_s = 180^{\circ}$, and $\theta_s$ decreases as the observer moves away
from that direction in either polar angle or azimuth.

All angles quoted in this work, for the sun, the sky and the observer, are
measured in air outside the module. Light reaching the coating is refracted
at the front glass, so the polar angles that enter
Eq.~\eqref{eq:thetascatter} are the internal ones,
$\theta_{\mathrm{in}} = \arcsin(\sin\theta/n_{\mathrm{med}})$; since the glass
and the encapsulant are plane-parallel, only $n_{\mathrm{med}}$ enters. An
external sweep from $0^{\circ}$ to $45^{\circ}$ therefore spans only
$0^{\circ}$ to $28^{\circ}$ inside EVA, and no external direction reaches
beyond the critical angle of about $42^{\circ}$. The angular Fresnel
transmission of the front surface is applied on the way in and on the way
out, normalized to its value at normal incidence.

Equations \eqref{eq:qtheta} and \eqref{eq:thetascatter} together frame the
angular question posed in Section~\ref{sec:intro}, and it is worth stating
the point precisely because it is often elided. The structure factor depends
only on the magnitude of the momentum transfer and not on its orientation, so
the correlated enhancement is isotropic and carries no preferred viewing
direction; this is what guarantees the absence of Bragg iridescence. But the
momentum transfer accessible to a given observer is itself set by the
geometry, so an observer who moves is not sampling a different orientation of
an anisotropic structure, but a different point on the isotropic curve
$S(q)$, and simultaneously a different point on the resonant phase function
$p(\theta)$. Since $S(q)$ varies by more than an order of magnitude even across
the backward scattering angles accessible through the front glass at the packing fractions of interest, the isotropy of
$S$ does not by itself guarantee an invariant perceived color. Whether the
color is stable is an empirical question about the specific recipe, and it is
answered in Section~\ref{sec:results}.

\subsection{From the packing to the perceived color}\label{sec:stack}

The coating is a film of finite thickness composed of one or more correlated
monolayers, deposited on the inner face of the front glass and embedded in the encapsulant above the photovoltaic cell. Propagation through
this stack is treated as a two-flux radiative-transfer problem of the
Kubelka--Munk type \citep{KubelkaMunk1931, Kubelka1948}, in which each layer
returns, transmits or absorbs the incident flux and the layers are coupled
incoherently to the underlying substrate.

The optical thickness of a single correlated monolayer follows from its
geometry alone. For spheres of diameter $d$ packed at volume fraction $\phi$,
a monolayer of thickness $d$ contains $6\phi/(\pi d^2)$ spheres per unit
area, each presenting a scattering cross-section $Q_{\mathrm{sca}}\pi d^2/4$,
so that
\begin{equation}
  \tau(\lambda) = \tfrac{3}{2}\,\phi\, Q_{\mathrm{sca}}(\lambda),
  \label{eq:tau}
\end{equation}
independent of sphere size, as it must be, since increasing $d$ at fixed
$\phi$ enlarges each cross-section and thins the areal number density in
equal measure. Weighting this optical thickness by the returned fraction
$W_b(\lambda)$ of the preceding section gives the layer reflection albedo
\begin{equation}
  r_b(\lambda) = 1 - \exp\!\big[-\tau(\lambda)\, W_b(\lambda)\big],
  \label{eq:rb}
\end{equation}
with the transmission
$t = (1 - r_b)\exp\!\big(-\tfrac{3}{2}\phi\, Q_{\mathrm{abs}}\big)$, so that a
transparent layer redistributes the flux and an absorbing one also removes
part of it on every pass. In the notation of Kubelka and Munk, the products
$\tau W_b$ and $\tfrac{3}{2}\phi Q_{\mathrm{abs}}$ take the place of $Sd$ and
$Kd$, the scattering and absorption coefficients multiplied by the layer
thickness: for an optically thin layer $r_b \to \tau W_b$, and the fraction
absorbed on one pass is $\tfrac{3}{2}\phi Q_{\mathrm{abs}}$. Rather than being
fitted, or converted from single-particle quantities with factors that depend
on the assumed angular distribution of the flux, both are fixed here by the
Mie cross-sections, the structure factor and the packing geometry. The absorption path is taken as a single normal
pass through the layer; doubling it, as for fully diffuse flux, changes the
colors of the absorbing recipes by $\Delta E_{00}$ of $1.1$ to $3.0$
(Supplementary Material). Equations~\eqref{eq:tau} and \eqref{eq:rb}
leave the model free of adjustable optical parameters: the packing fraction
enters the reflectance twice, once through the structure factor and once
through the layer optical thickness, and both dependences are fixed by the
geometry rather than fitted. The spectral selectivity of the coating resides
in the product $\tau W_b$. For a representative blue recipe the layer albedo exceeds $0.6$ at the wavelength where the backscattering condition coincides with the maximum of $S(q)$, and falls below $0.01$ in the red, so a single
monolayer is a strong reflector within its band and effectively transparent
outside it. A stack of $N$ such layers over the substrate is then assembled
by the incoherent two-flux recursion, and the achromatic Fresnel reflection
of about four percent at the air-glass interface is added by one further
application of the same combination, yielding the total spectral reflectance
$R(\lambda)$ of the module. Both steps are given in the Supplementary
Material. Only the reflected flux is evaluated, since it alone sets the
perceived color; the spectral transmittance into the cell, and with it the
photocurrent retained under each recipe, lies outside the scope of this work.
Qualitatively, a transparent coating costs photocurrent only through the light
it returns, whereas an absorbing one also removes the fraction
$\tfrac{3}{2}\phi Q_{\mathrm{abs}}$ on every pass.

Because the recursion accumulates multiply scattered flux, increasing $N$
raises the luminous reflectance and progressively decouples the returned
light from the single-scattering geometry. Layer number therefore acts
primarily on the lightness of the coating while the diameter selects its hue,
and whether the decoupling improves or degrades the angular stability of the
perceived color is a separate question, settled numerically in
Section~\ref{sec:results}.

The two-flux formulation is chosen for its transparency and low cost: it follows the dependence of the returned flux on every design parameter while carrying no fitted optical parameter, so the reflectances and colors reported below follow from the particle and stack geometry alone. As a two-stream approximation it does not resolve the angular distribution of light within the stack, and the results that depend most directly on the treatment of multiple scattering, those concerning the layer number, are identified as such where they arise.

A façade is not illuminated by a single collimated beam, and since the
central claim of an amorphous colorant is that its color survives the daily
variation in illumination, the incident light is modeled explicitly rather
than assumed away. Four levels of description are used, in increasing order
of realism, all sharing the geometry of Fig.~\ref{fig:schematic}(b). A
\emph{single directional beam} isolates the intrinsic angular response and is
the configuration in which the target colors are defined, at normal incidence
and normal observation. An \emph{isotropic sky}, the fully overcast limit, has
uniform radiance over the external hemisphere; refracted into the encapsulant
it fills a cone up to the critical angle, so that for a normal observer every
scattering angle from about $138^{\circ}$ to $180^{\circ}$ contributes
simultaneously to a single perceived color. It tests whether the correlated
angular response survives this averaging, and as Section~\ref{sec:sky} shows,
the answer differs sharply between blue and the other tones. A
\emph{clear sky} superposes the two, with the direct share taken from the
clear-sky model introduced below at a solar polar angle of $30^{\circ}$,
$0.88$, close to the ratio of $0.875$ between the direct and global spectra
of \cite{ASTMG173} across the visible, and a hazier sky with a direct share
of $0.70$ examined as a sensitivity case. Finally, the fixed weighting is
relaxed so that the diffuse share follows the sun: as the solar elevation
falls the optical air mass grows, the beam is attenuated more strongly, and
the diffuse fraction rises. We evaluate it with the broadband clear-sky
model of Bird and Hulstrom \cite{BirdHulstrom1981}, as implemented in pvlib
\cite{pvlib2023}, for the atmosphere that defines the AM1.5G spectrum \cite{ASTMG173}: an aerosol
optical depth of $0.084$ at $500$\,nm, $1.42$\,cm of precipitable water and
$0.34$\,cm of ozone, with the aerosol optical depth at $380$\,nm following
from the {\AA}ngstr\"{o}m law with the classical exponent of $1.3$
\cite{Angstrom1964}. This gives diffuse fractions of $0.111$, $0.129$ and
$0.179$ at solar polar angles of $20^{\circ}$, $40^{\circ}$ and $60^{\circ}$,
so that a single sweep of the solar position changes both the geometry of the
direct beam and its weight relative to the sky. These three positions are
used throughout to represent midday, mid-afternoon and low-sun conditions. In
every case the reflectance under a composite sky is the irradiance-weighted
sum of the spectral reflectances for its individual incident directions, so
the layer coupling, the substrate and the colorimetry are common to all of
them and any color change observed between them is attributable to the
illumination alone.

The spectral reflectance is finally converted to a perceived color. Color is
computed from the CIE~1931 $2^{\circ}$ standard colorimetric observer under
the AM1.5G global spectrum \citep{CIE15, ASTMG173}, so that the predicted
appearance corresponds to a module viewed under standard terrestrial solar
illumination. Chromaticity is reported in the CIE~1976 $u'v'$ uniform
chromaticity scale rather than in the 1931 $xy$ diagram, since equal
distances in the latter do not correspond to equal perceived differences, and
for the same reason the vividness of a color is quantified by the CIELAB
chroma $C^{*}_{ab}$ rather than by a distance in the chromaticity plane. The
reference white is the perfect diffuser under AM1.5G rather than a standard
daylight illuminant, so that the white point is consistent with the
illumination under which the colors themselves are computed. The same
colorimetric treatment, applied to emissive display primaries rather than to
a reflective coating, is described in \cite{Genc2017Rec2100}.

Differences between colors, which are central to every comparison made here,
are evaluated with the CIEDE2000 formula \cite{LuoCuiRigg2001} in the
implementation of \cite{Sharma2005CIEDE2000}, with the parametric factors
set to the CIE reference values $k_L = k_C = k_H = 1$. Values other than
unity apply to specific industrial viewing situations, such as $k_L = 2$ in
textile assessment, and no validated parametric set exists for architectural
surfaces viewed at a distance, so the reference values are retained
throughout and $\Delta E_{00}$ is used here as a relative measure of color
separation rather than as an absolute tolerance. Two quantities are built
from it. The first is the distance from a computed color to one of the target
colors, used to decide whether a given recipe reaches a required façade tone.
The second is the angular drift of a recipe, defined as the largest
$\Delta E_{00}$ between any two viewing configurations within a specified
range of illumination and observation angles, which measures how far the
appearance of a coating changes as the sun and the observer move. As a guide
to interpretation, a difference of $\Delta E_{00} \approx 1$ is at the
threshold of perceptibility for adjacent samples under reference conditions,
and differences above about $5$ read as distinct colors. The colorimetric
transformations, the color swatches rendered in sRGB for display, and the
treatment of colors falling outside the sRGB gamut are all set out in the
Supplementary Material.

\section{Results and Discussion}\label{sec:results}

All results below are computed with the framework of
Section~\ref{sec:model}, whose numerical implementation is verified in the
Supplementary Material. Unless a parameter is itself being swept, the
reference configuration is a coating of $N = 3$ correlated monolayers at a
packing fraction of $\phi = 0.64$, embedded in EVA over a TopCon substrate
and evaluated under AM1.5G at normal incidence and normal observation. Since
the model contains no fitted optical parameter, the reflectances and the
comparisons between them follow from the particle and stack geometry alone.

\subsection{The reachable palette and the target colors}\label{sec:palette}

\begin{figure*}[t]
  \centering
  \includegraphics[width=\textwidth]{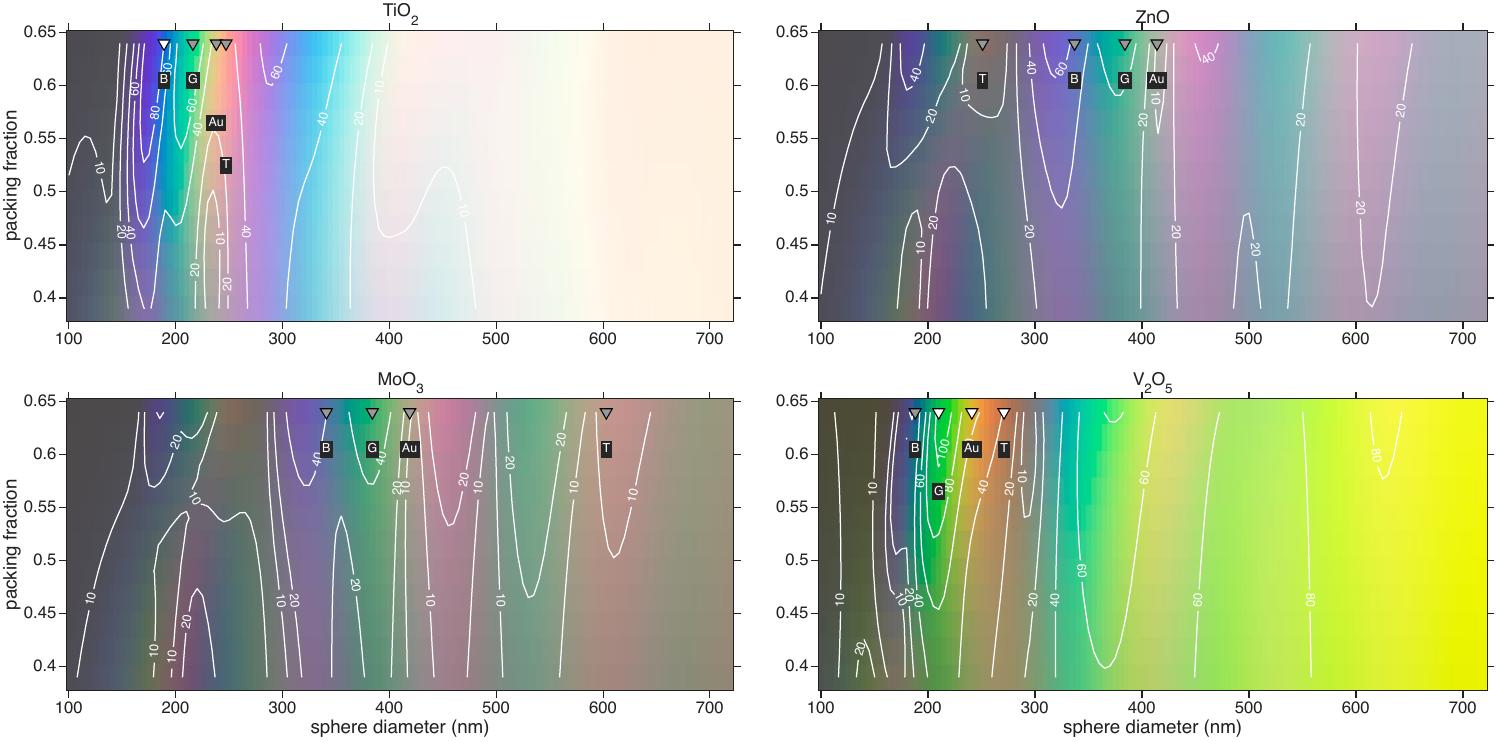}
  \caption{Rendered color of a photonic-glass coating as a function of sphere
  diameter and packing fraction, one panel per candidate oxide, for $N = 3$
  layers in EVA over a TopCon substrate at normal incidence and normal
  observation under AM1.5G. White contours give the CIELAB chroma
  $C^{*}_{ab}$. Triangles at $\phi = 0.64$ mark each oxide's closest match to
  the four target colors (B, G, Au and T for blue, green, gold and
  terracotta); white triangles mark the best match among all four oxides,
  grey triangles the matches that are not the best.
  The transparent TiO$_2$ covers the cool half of the hue circle and renders
  its warm half pale, whereas the blue-absorbing V$_2$O$_5$ renders green,
  amber and terracotta deeply but its blue dark.}
  \label{fig:sizemap}
\end{figure*}

The primary control over the perceived color is the sphere diameter, since it
sets the spectral position of the Mie resonance that the short-range order
then reinforces. Figure~\ref{fig:sizemap} maps the rendered color and the
CIELAB chroma across diameter and packing fraction for each of the four
oxides, from $100$ to $720$\,nm. Three features organize everything that
follows.

First, the transparent and the absorbing high-index oxides cover different
halves of the hue circle. In TiO$_2$ the hue sweeps from violet through a
vivid blue and a blue-green to pale sand, pink and lilac as the diameter
passes from about $170$ to $300$\,nm, and beyond about $350$\,nm the color
fades towards white as overlapping resonances fill the visible band. The warm
part of this sequence is pale: TiO$_2$ returns broadband light on either side
of a resonance placed at the band edge, so its warm tones are light and of low
chroma. V$_2$O$_5$, which absorbs below about $500$\,nm, removes that
short-wavelength return and turns the same part of the sequence into a deep
green near $210$\,nm, amber and terracotta near $250$ to $280$\,nm and, at
large diameters where the resonances wash out, a yellow that owes more to
absorption than to scattering. Its blue, by contrast, is dark.

Second, the reachable chroma differs sharply between materials. Over the
mapped range TiO$_2$ spans 0.4 to
99.7, V$_2$O$_5$
0.8 to
108.3, ZnO
1.0 to
68.1 and MoO$_3$
1.6 to
56.4. Among the transparent oxides the
ordering follows the index contrast of Eq.~\eqref{eq:relindex} directly: in
EVA a TiO$_2$ sphere presents a contrast of about $1.7$ against the medium,
whereas ZnO and MoO$_3$ present only about $1.3$, and their palettes are
correspondingly muted. V$_2$O$_5$ exceeds TiO$_2$ because absorption deepens
the color it scatters. All four oxides also pass through a perceptually
neutral region within the mapped range.

Third, the packing fraction acts as a secondary lever whose sign depends on
the diameter. Near a resonance a denser packing sharpens the structure-factor
peak and raises the chroma, the mechanism by which correlated disorder
produces saturated color in the first place
\citep{Schertel2019StructuralColors}, whereas off resonance it reinforces a
spectral region where the single-particle response is already flat and the
color washes out. Packing fraction is therefore a refinement to be applied
once the diameter has selected the hue, and we hold it at the random
close-packed value $\phi = 0.64$ throughout.

From the laminated palette we fix four target colors, a blue, a green, a gold
and a terracotta, the families that recur in façade practice
(Section~\ref{sec:intro}). A target named by its authors is open to the
objection that it is not what its name says, so we choose the targets by an
external criterion. Candidate colors of the laminated palette, across
diameters and layer numbers, were rendered in sRGB and named against the
color survey of \cite{XKCDColorSurvey2010}, in which more than five million
color namings from $222{,}500$ sessions were reduced to a published list of
$949$ named colors. For each family we took a color whose closest name among
the basic and most common color names of the survey is the family's own, with
the largest margin to any name outside the family. The four targets,
\texttt{\#3067ea}, \texttt{\#00c150}, \texttt{\#e4be28} and
\texttt{\#c7714e}, have as their nearest survey names
\emph{bright blue}, \emph{azul}, \emph{cerulean blue};
\emph{shamrock green}, \emph{shamrock}, \emph{cool green};
\emph{gold}, \emph{dark yellow}, \emph{golden}; and
\emph{adobe}, \emph{terracota}, \emph{terracotta},
and each lies at least 8.0
CIEDE2000 units closer to its own family name than to any other
(Supplementary Material). In CIELAB under AM1.5G they sit at
$L^{*} = 47.2$, $C^{*}_{ab} = 84.3$,
$h_{ab} = 289^{\circ}$ (blue),
$65.2$, $105.6$, $161^{\circ}$ (green),
$78.3$, $69.0$, $92^{\circ}$ (gold) and
$56.6$, $40.9$, $46^{\circ}$ (terracotta).
They are more saturated than the muted tones of most façades, which is
deliberate: muting a reached color is straightforward, by lowering the
packing fraction or the coverage, whereas reaching an unambiguous hue is the
harder test. A saturated red remains out of reach for every material and every
combination of diameter, packing fraction and layer number examined,
consistent with the known constraint on correlated disordered scattering
\citep{Magkiriadou2014RedAbsence, Jacucci2020Limitations}; terracotta is the
reddest of the four families that a dielectric photonic glass can render.

The layer number deserves a word, since it is the one geometric parameter
not fixed by the color target. It is set here by the targets themselves. In
the best recipes of Section~\ref{sec:materialswap}, the gold is named gold
only from $N = 2$ upwards, because a single layer is too dark, and the
terracotta turns orange from $N = 6$ upwards, as the accumulating scattered
light lightens it; all four keep their names between $N = 2$ and $N = 5$, and
the largest distance of the four from their targets is smallest at $N = 3$,
2.5 against
5.0 at $N = 4$ and
5.4 at $N = 2$. We therefore
adopt $N = 3$ throughout. It is also tolerant of the layer-count variation that
a sprayed coating carries: one layer more or fewer moves the four colors by
$\Delta E_{00}$ of 1.0
to 4.6, without changing
their names, and for the TiO$_2$ blue $N = 3$ is the chroma maximum itself.

It is worth noting what these colors are not. All four have a luminous
lightness above that of the module they sit on: the uncolored TopCon cell seen
through the front glass renders at $L^{*} = 31.8$, or
$24.3$ for the physics-based textured
cell of Section~\ref{sec:materials}, whereas the four best recipes sit between
$L^{*} = 45.0$ and
75.2. A purely absorptive colorant in front
of the cell can only remove flux from what the cell already returns and is
therefore bounded above by that $L^{*}$; scattering pigments and printed
ceramic inks can exceed it, but they do so by scattering, the mechanism used
here. Because the color is produced largely by redirecting rather than
absorbing light, its cost to the cell is set by how much flux the coating
returns and, for an absorbing oxide, absorbs, a quantity we address together
with fabricated modules in ongoing experimental work.

\subsection{Which oxide reaches which target, and the effect of lamination}\label{sec:materialswap}

\begin{figure*}[t]
  \centering
  \includegraphics[width=0.9\textwidth]{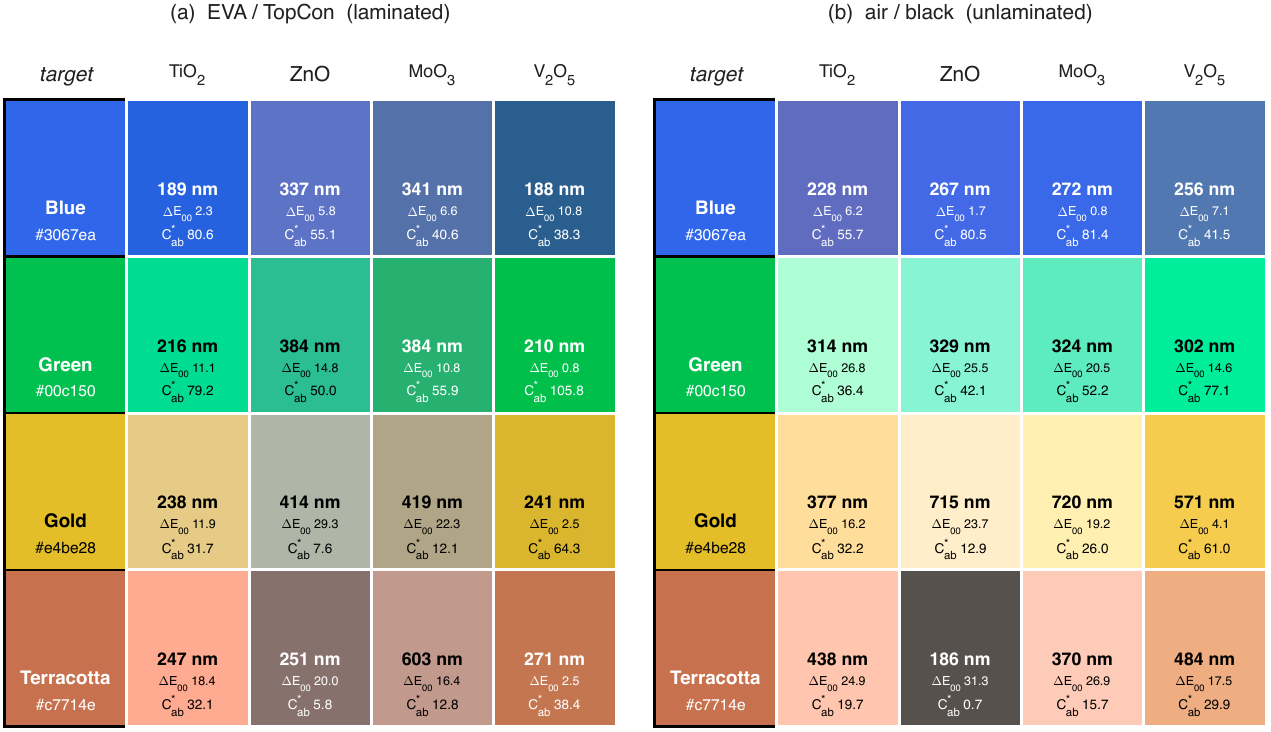}
  \caption{Closest reachable color to each of the four target colors, per
  oxide, at $N = 3$ and $\phi = 0.64$ with the diameter scanned from $120$ to
  $720$\,nm. The leftmost column shows the targets. Each cell gives the
  matched diameter, the CIEDE2000 distance to the target and the CIELAB
  chroma of the match. (a)~EVA over TopCon, the laminated stack in which the
  targets are defined. (b)~The same search in air over a black substrate, as a
  coating would be inspected before lamination.}
  \label{fig:match}
\end{figure*}

Whether a manufacturer is free to choose the scattering material depends on
whether the same color can be produced from a different oxide.
Figure~\ref{fig:match} answers this by scanning the diameter of each material
from $120$ to $720$\,nm and reporting the closest approach to each target.
Since the targets are fixed colors rather than the output of any one material,
every oxide is measured against them on the same footing.

The answer is that the palette divides between the two high-contrast oxides.
TiO$_2$ reaches the blue, at $189$\,nm and
$\Delta E_{00} = 2.3$, but none of the others: its closest
green, at $216$\,nm, lies $11.1$ away and
is a turquoise, its closest gold $11.9$ away is a sand, and its
closest terracotta $18.4$ away is a light salmon, by the
same naming criterion.
V$_2$O$_5$ reaches the other three, the green at $210$\,nm
and $\Delta E_{00} = 0.8$, the gold at
$241$\,nm and $2.5$, and the terracotta at
$271$\,nm and $2.5$, but not
the blue, which it renders $10.8$ away as a darker denim blue,
because the absorption below $500$\,nm removes much of the blue return itself.
The same absorption is what makes the other three reachable: suppressing the
short-wavelength return deepens a warm tone that a transparent scatterer can
only render pale, and narrows the green. These four recipes, TiO$_2$ at
$189$\,nm and V$_2$O$_5$ at $210$, $241$ and
$271$\,nm, are the reference recipes for the rest of the paper.

This result reframes the role of the index ceiling. A transparent oxide at the
ceiling reaches the cool targets by resonance, but its warm palette is pale,
and the only way past that within the oxides is to accept absorption in the
blue, which brings warm tones and green within reach at a cost in the band
the cell converts. The low-contrast oxides illustrate the other side.
ZnO and MoO$_3$ render muted blues, $5.8$ and
$6.6$ from the target, and only through their second-order
resonance at $337$ and $341$\,nm; MoO$_3$ renders
a jade green $10.8$ away and ZnO a blue-green
$14.8$ away; and both fail at the warm end, their closest gold
and terracotta being greys, stone and pinkish greys of chroma
$5.8$ to
$12.8$. Contrast decides
whether a color exists at all; absorption decides which colors a high-contrast
oxide can render.

The second panel of Fig.~\ref{fig:match} isolates the effect of lamination.
Searching again in air over a black substrate, the state in which a freshly
deposited layer would be inspected, the closest approaches lie at different
diameters from the laminated recipes, for example $228$
instead of $189$\,nm for the TiO$_2$ blue and
$571$ instead of $241$\,nm for the
V$_2$O$_5$ gold, and the green and terracotta are not reached in air at all,
the closest being $14.6$ and
$17.5$ away. Applying the reference
recipes themselves rather than re-optimizing makes the point more sharply
still. The same four coatings, laminated, render as the targets; surrounded by
air over black they shift by $\Delta E_{00} = 17.6$,
$39.2$, $43.5$ and $45.1$, because the
contrast in air is high enough to move the structure-factor maximum to the
short-wavelength end of the visible; the whole diameter and packing map in
air is displaced accordingly (Fig.~S4). Three of the four move by more than the
difference between two unrelated façade colors, and the blue by far more than
a distinct-color step.

This is a practical constraint rather than a curiosity. A coating cannot be
developed, inspected and color-matched in the unlaminated state and then
encapsulated, because the encapsulant does not merely tint the result, it
relocates the resonance. The constraint aligns with the manufacturing
argument of Section~\ref{sec:intro}: since glass-side deposition delivers the
coating into the laminate in any case, the design must be carried out for the
laminated stack from the outset, which is what we do throughout.

The contrast with the dilute silicon regime is instructive. Song et
al.~\cite{Song2026SubMonolayer} report that their reflectance peaks remain
spectrally stationary when the sub-monolayer is integrated into the module. A
dilute resonator colors at its own Mie resonance, whose wavelength is set
chiefly by the particle index and size, so a change of surroundings
attenuates it without moving it. A correlated packing colors where the
structure-factor peak falls, and since the momentum transfer scales with the
index of the surrounding medium, that wavelength moves with the medium.
Sensitivity to lamination is therefore intrinsic to the correlated regime
rather than a shortcoming of the oxides. It can, however, be removed by
architecture. In the photonic pigments of \cite{Li2025Nexus} the medium
around each sphere is an acrylate cured inside the grain, so the encapsulant
surrounds the grain rather than the sphere and no longer enters the contrast.
The same step applied to an oxide packing, infiltration with a polymer of
fixed index followed by grinding, would fix the color before lamination, at
the price of giving up the encapsulant as a design variable.

By contrast, the choice of cell technology beneath the coating is almost
irrelevant. Placing the reference recipes on EVA over PERC rather than TopCon
changes their color by $\Delta E_{00}$ of
0.03 to
0.32, well below the threshold of
perceptibility; the full comparison is given in Fig.~S7. The medium matters
and the substrate does not, which is a convenient asymmetry: a single coating
recipe can be specified for a module line without regard to which silicon
technology it will sit on. The same insensitivity does not extend to the
encapsulant itself. Replacing EVA by POE or PVB moves the green reference by
only 1.13 and 0.43, but
replacing it by the higher-index TPU moves it by 7.09,
of which retuning the diameter to 204\,nm recovers
all but 0.90; the comparison across all four
encapsulants is given in Figs.~S8 and S9.

\subsection{Behavior under realistic sky illumination}\label{sec:sky}

\begin{figure*}[t]
  \centering
  \includegraphics[width=0.8\textwidth]{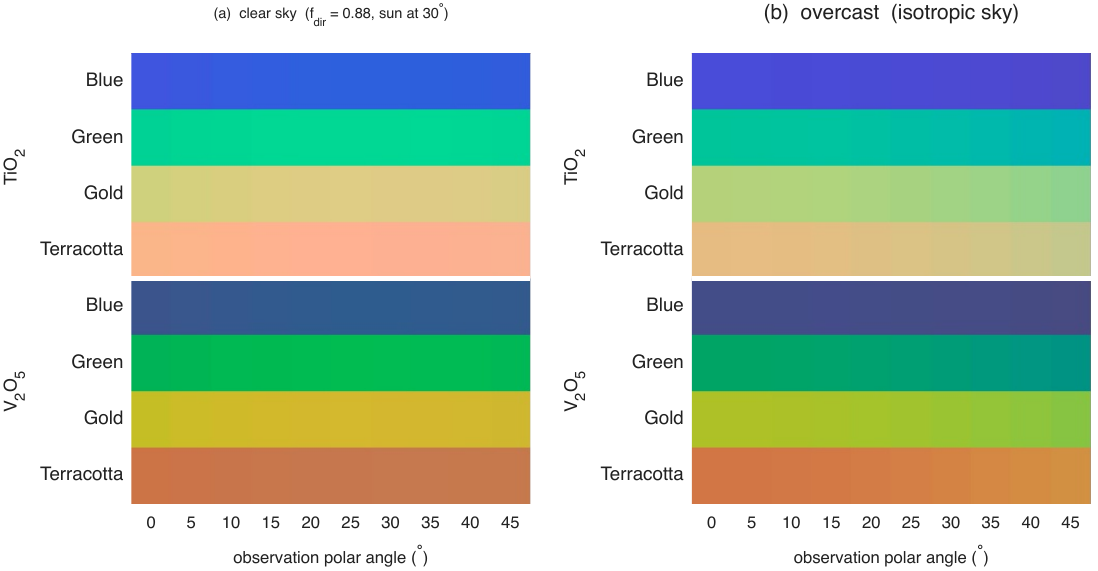}
  \caption{Rendered color against observation polar angle under two sky
  models, for the matched recipes of the four targets in TiO$_2$ (upper block)
  and V$_2$O$_5$ (lower block). (a)~Clear sky, a direct beam at a solar polar
  angle of $30^{\circ}$ carrying $88\%$ of the flux
  superposed on isotropic skylight. (b)~Fully overcast, isotropic sky. Blue is
  almost invariant under the overcast sky, drifting by
  $\Delta E_{00} = 2.2$ for TiO$_2$ and
  $2.3$ for V$_2$O$_5$, whereas the other recipes
  drift by $9.2$ to
  $15.9$. All angles are
  measured in air.}
  \label{fig:sky}
\end{figure*}

The colors of the preceding sections were computed for a single collimated
beam at normal incidence, which is the condition under which a coating would
be characterized in a laboratory but not the condition under which a façade
is seen. Figure~\ref{fig:sky} replaces that idealization with two sky models
and follows the color as the observer moves from normal to $45^{\circ}$ off
normal, for the matched recipes in the transparent and in the absorbing
high-contrast oxide.

Under a clear sky, a direct beam at a solar polar angle of $30^{\circ}$
carrying $88\%$ of the flux together with isotropic
skylight, the reference recipes drift by $\Delta E_{00}$ of
$4.8$ (TiO$_2$ blue), $3.4$,
$5.2$ and $2.6$
(V$_2$O$_5$ green, gold and terracotta) as the observer moves from normal to
$45^{\circ}$, and the remaining TiO$_2$ and V$_2$O$_5$ matches by
$2.8$ to
$6.7$. Since a difference of about five is read as a
distinct color, an observer moving in the plane of incidence under a fixed sun
sees the colors change by about one distinct-color step at most; the
remainder of this section and the next identify the conditions under which
they change more.

The fully overcast limit, in which the radiance is uniform over the sky,
separates blue from the other tones. Blue becomes almost perfectly invariant,
drifting by only $2.2$ in TiO$_2$ and
$2.3$ in V$_2$O$_5$. The other recipes move in the
opposite direction, the V$_2$O$_5$ green, gold and terracotta drifting by
$14.4$, $9.2$ and
$13.3$, between 1.8 and 5.1 times their clear-sky values.
Averaging over all incident directions therefore does not simply wash out the
angular dependence. The reason is that the average is taken over a correlated
angular response that is itself far from flat, and that the range it samples
widens as the observer tilts: refracted, the overcast sky supplies scattering
angles from $138^{\circ}$ to $180^{\circ}$ to a normal observer but from about
$110^{\circ}$ to $180^{\circ}$ to one at $45^{\circ}$. Over this range the
coefficient of variation of $p(\theta)\,S(q(\theta))$, averaged over the visible
band, is lowest for the blue recipes, $0.53$ in TiO$_2$ and
$0.49$ in V$_2$O$_5$, against
$0.62$ to
$0.82$ for the other recipes, so every
observation direction lands on much the same average for blue and not for the
others. Absorption plays no part in this: removing it from the layer
transmission changes the overcast drift of the V$_2$O$_5$ recipes by at most
1.4 units.

\begin{figure*}[t]
  \centering
  \includegraphics[width=0.8\textwidth]{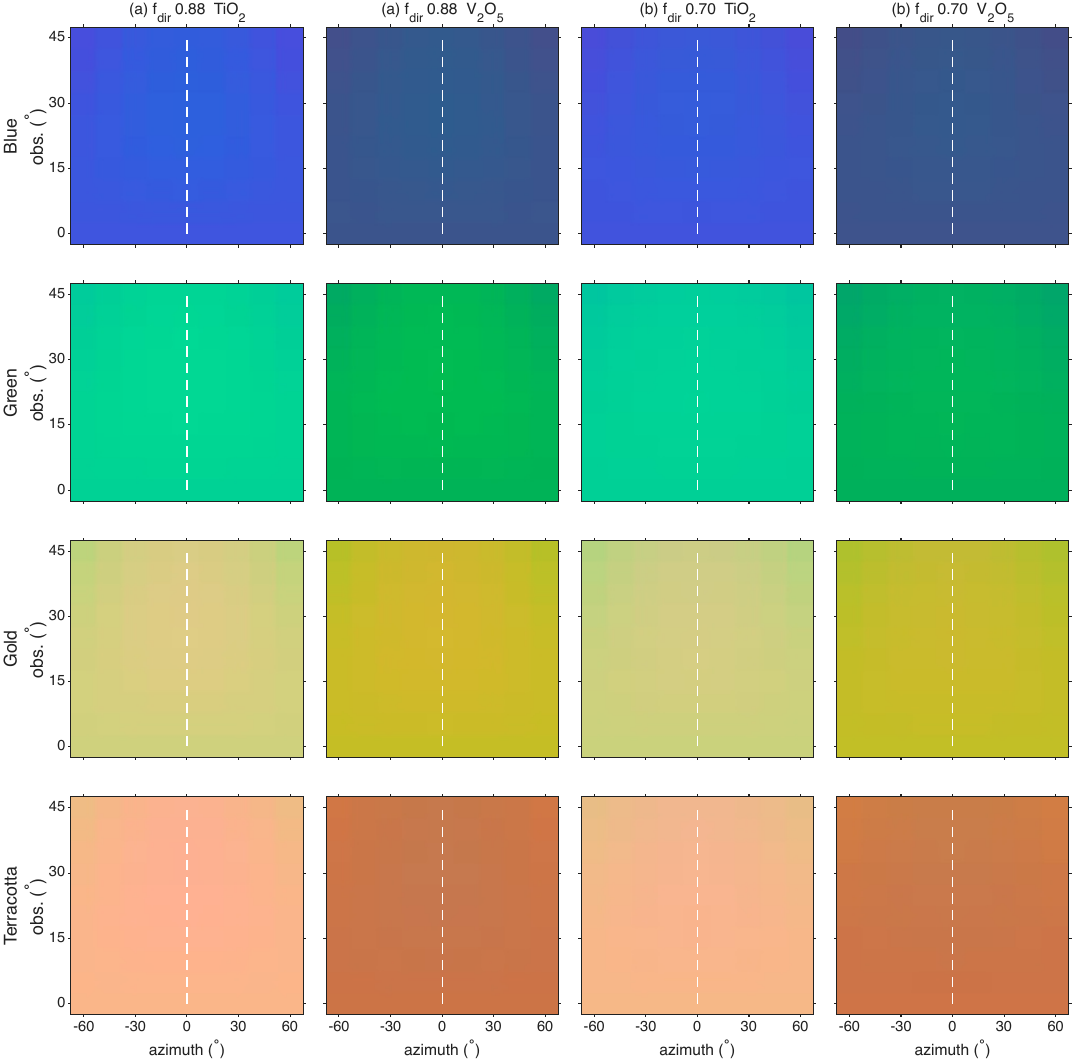}
  \caption{Rendered color under mixed illumination, a direct beam at a solar
  polar angle of $30^{\circ}$ superposed on isotropic skylight, swept over
  observation polar angle and azimuth (in air). (a)~Clear sky with a direct
  fraction of $0.88$. (b)~Hazier sky with a direct
  fraction of $0.70$. Columns alternate TiO$_2$ and V$_2$O$_5$ within each
  panel. Adding diffuse light changes the drift by at most
  1.9 CIEDE2000 units.}
  \label{fig:mixazi}
\end{figure*}

Introducing a realistic azimuthal spread and a hazier sky does not change this
picture. Figure~\ref{fig:mixazi} sweeps the observer over both polar angle and
azimuth under the same mixed illumination; for the reference recipes the drift
is $7.3$, $7.2$,
$9.6$ and $3.0$, and
reducing the direct fraction from $0.88$ to $0.70$ changes
it by at most 1.9
CIEDE2000 units for any recipe, lowering it for the blue and gold recipes and
raising it for the green and for the V$_2$O$_5$ terracotta. A hazy sky does not convert an
angle-dependent coating into an angle-independent one. The corresponding grids
at solar polar angles of $40^{\circ}$ and $60^{\circ}$ are given in Fig.~S6.

\subsection{Angular response, the role of thickness and the façade geometry}\label{sec:angular}

\begin{figure*}[t]
  \centering
  \includegraphics[width=0.8\textwidth]{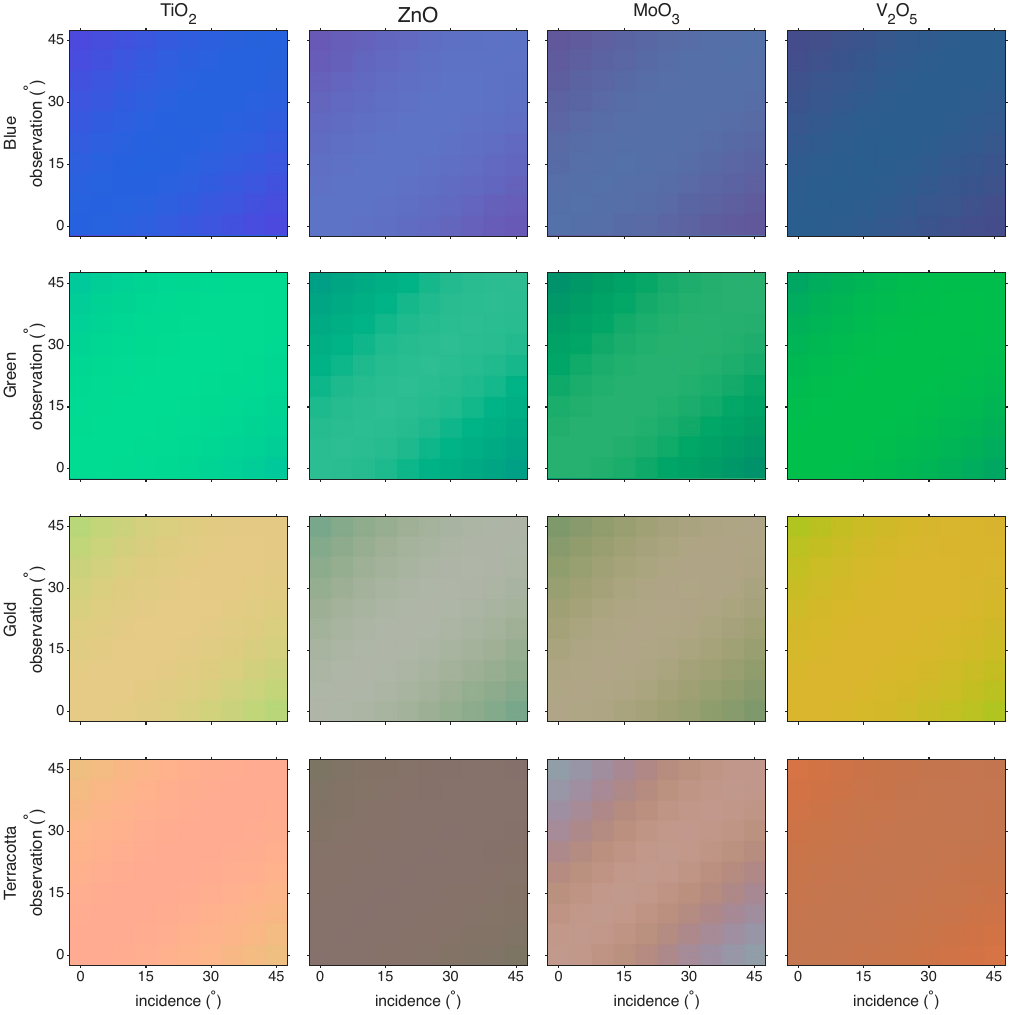}
  \caption{Rendered color over the two-dimensional illumination and
  observation grid, each angle swept from $0^{\circ}$ to $45^{\circ}$ in air in
  the plane of incidence, at $N = 3$ and $\phi = 0.64$ in EVA over TopCon.
  Columns are the four oxides at their matched diameters, rows the four target
  colors. The largest CIEDE2000 difference within each panel ranges from
  $4.5$ to $26.5$
  across the sixteen combinations.}
  \label{fig:obsinc}
\end{figure*}

\begin{figure*}[t]
  \centering
  \includegraphics[width=0.9\textwidth]{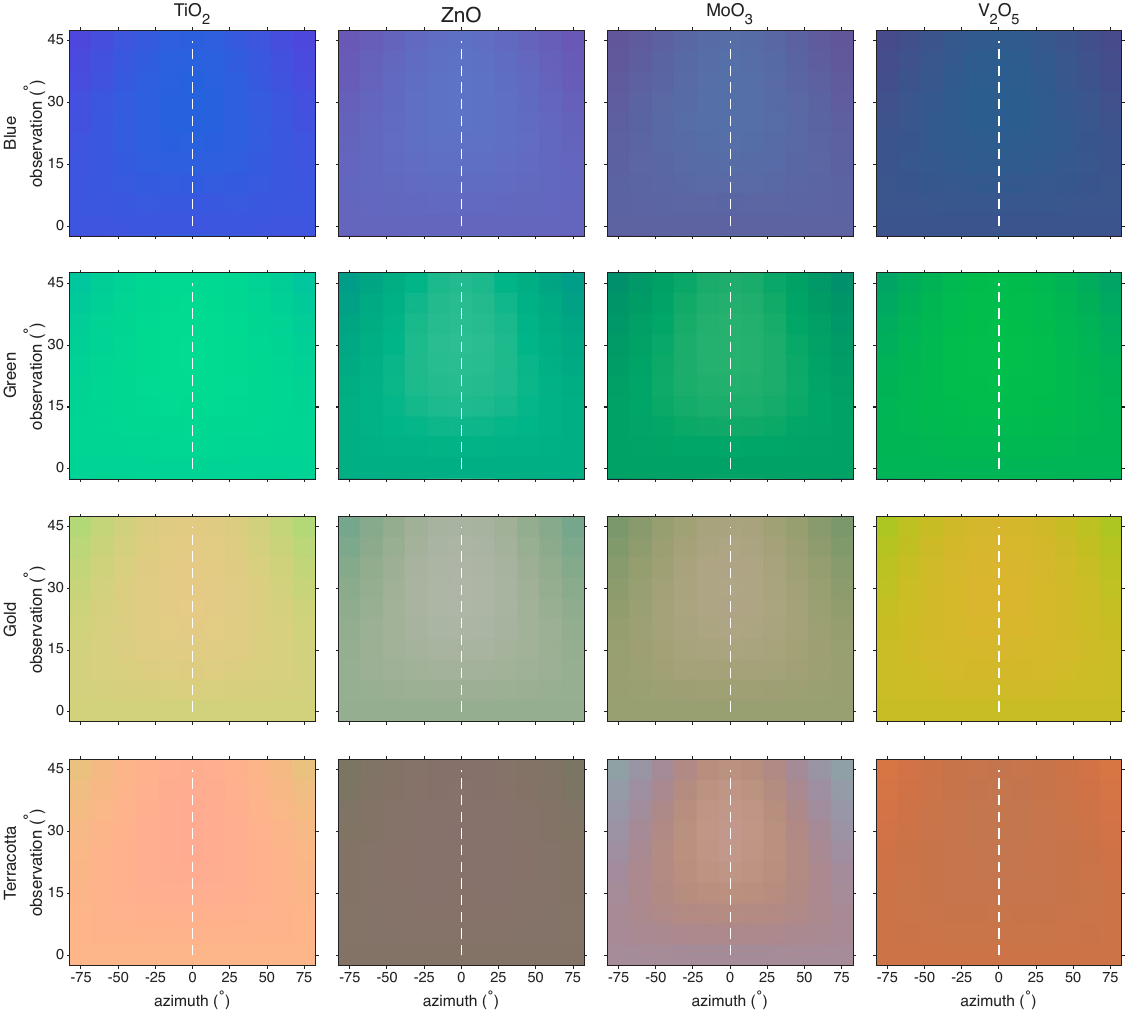}
  \caption{Rendered color over observation polar angle and azimuth at a fixed
  illumination polar angle of $30^{\circ}$ (in air). Azimuth is measured from
  the plane of incidence, so the dashed line at $\varphi_o = 0$ contains the
  retroreflection direction, reached at an observation angle of $30^{\circ}$.
  Columns are the four oxides at their matched diameters, rows the four
  targets. The panels are symmetric about $\varphi_o = 0$, as a spherical scatterer
  and an isotropic structure factor require, and drift across the panel spans
  $4.5$ to $29.2$.}
  \label{fig:obsazi}
\end{figure*}

The central claim made for amorphous photonic materials is that their color
is angle-independent. Section~\ref{sec:phase} established that the isotropy
of the structure factor does not by itself secure this, since a moving
observer samples a different point on $S(q)$ and on the phase function
simultaneously. Figures~\ref{fig:obsinc} and \ref{fig:obsazi} test the claim
across the two angular degrees of freedom available to a façade, for all four
materials at their matched diameters.

Figure~\ref{fig:obsinc} sweeps illumination and observation independently
from $0^{\circ}$ to $45^{\circ}$ in the plane of incidence, that is, from
$0^{\circ}$ to $28^{\circ}$ inside the encapsulant. No combination of material
and target is invariant. For the reference recipes the drifts are
$9.9$ (blue), $10.1$ (green),
$13.8$ (gold) and $4.5$
(terracotta), and across all sixteen combinations the range is
$4.5$ to
$26.5$. What the disorder delivers is the
absence of the abrupt hue reversals that a Bragg stack would show: the color
changes smoothly over the whole grid. Compared with the observer sweep of
Fig.~\ref{fig:sky}, the drift here is larger because the illumination angle
also changes; the color responds to where the sun is more than to where the
observer stands.

Figure~\ref{fig:obsazi} adds the azimuthal degree of freedom, sweeping the
observer through $\pm 75^{\circ}$ out of the plane of incidence at a fixed
illumination angle of $30^{\circ}$. The magnitudes are comparable to the
in-plane case, $9.4$ to $19.2$ for
TiO$_2$ and $4.5$ to $14.9$ for
V$_2$O$_5$, and the panels are symmetric about $\varphi_o = 0$. This symmetry is
built into the description rather than tested by it: a spherical scatterer and
an isotropic $S(q)$ make the response depend on azimuth only through the
scattering angle of Eq.~\eqref{eq:thetascatter}. It is nevertheless the
property that distinguishes a photonic glass from an ordered structure, whose
lattice planes would break it, and what it does not do is make the color
constant. The corresponding grids at illumination angles of $40^{\circ}$ and
$60^{\circ}$ are given in Fig.~S5.

The V$_2$O$_5$ terracotta is the exception in both grids, drifting by only
$4.5$ and $4.5$, against
$17.6$ and $19.2$ for the closest TiO$_2$
terracotta. Its color is set largely by the absorption edge of V$_2$O$_5$,
which does not move with the scattering angle, rather than by a resonance,
which does; the other V$_2$O$_5$ recipes, whose colors are carried by
resonances, drift as much as their TiO$_2$ counterparts. Low contrast offers
no shelter either: the low-contrast matches drift by
$10.3$ to
$29.2$ across the two grids, and those that are grey at
normal viewing gain color at oblique geometry. Angular constancy is only
meaningful for a coating that has the target color to keep, which is why the
comparison that matters is between recipes that reach it.

\begin{figure*}[t]
  \centering
  \includegraphics[width=0.7\textwidth]{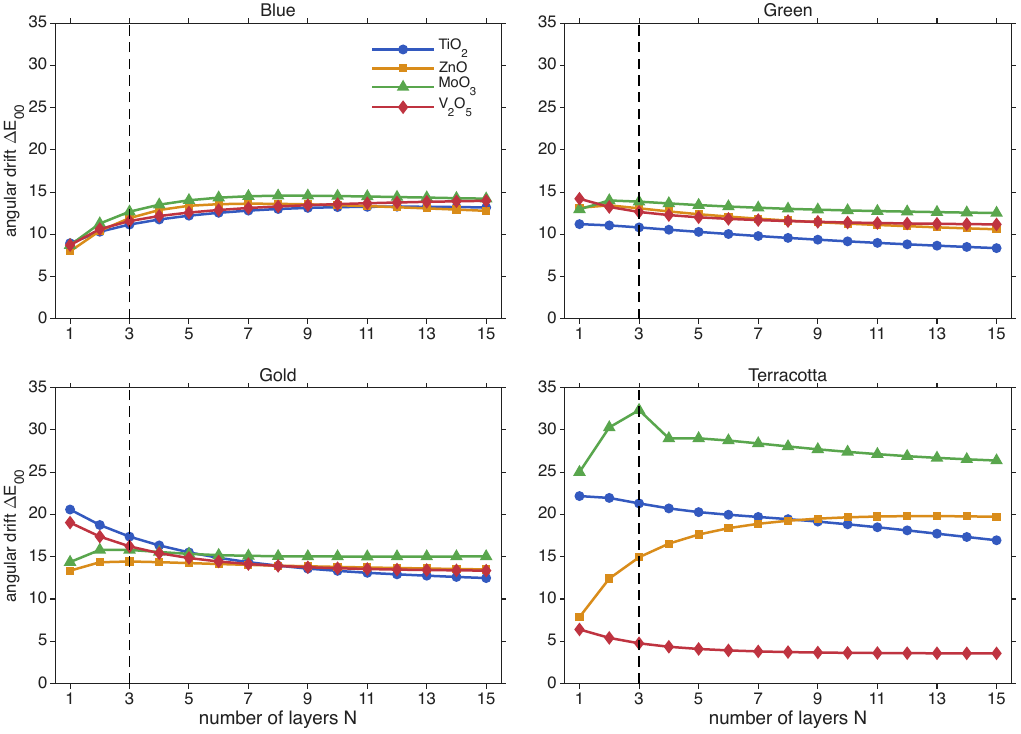}
  \caption{Angular drift against layer number, defined as the largest
  CIEDE2000 difference between any two illumination angles in the range
  $0^{\circ}$ to $50^{\circ}$ (in air) at normal observation, for $N = 1$ to
  $15$ at $\phi = 0.64$ in EVA over TopCon. One panel per target, one curve per
  oxide at its matched diameter; the dashed line marks the reference $N = 3$.
  The blue drift grows over the first few layers in every oxide and then
  levels off, the V$_2$O$_5$ green, gold and terracotta improve slowly, and the V$_2$O$_5$ terracotta is the most stable
  recipe at every $N$.}
  \label{fig:nangle}
\end{figure*}

Thickness is the obvious candidate remedy. The two-flux recursion of
Section~\ref{sec:stack} accumulates multiply scattered flux as layers are
added, and multiple scattering is known to reduce the angular dependence of
structural color in disordered media \citep{Hwang2020AngleIndependent}, so a
thicker coating might be expected to forget the single-scattering geometry.
Figure~\ref{fig:nangle} tests this over $N = 1$ to $15$ at normal observation,
with the illumination swept from $0^{\circ}$ to $50^{\circ}$. Thickness is a
lever, but a weak and two-sided one. The blue drift is larger at $N = 15$ than at
$N = 1$ in every oxide, although not monotonically in between, in TiO$_2$ from $9.0$ at $N = 1$ to
$13.2$ at $N = 15$, whereas the V$_2$O$_5$ green, gold and
terracotta fall slowly, from $14.2$ to
$11.2$, from $19.0$ to
$13.4$ and from $6.4$ to
$3.6$. At the reference $N = 3$ the four recipes
drift by $11.2$, $12.7$,
$16.2$ and $4.8$, within
1.8 units of their values at
$N = 6$, twice the reference thickness, so the thin coating that the color targets call for costs little in
angular stability. Thickness is best used to redistribute stability between
tones rather than as a remedy on its own; it is also the result that depends
most directly on the treatment of multiple scattering within the stack, and an
angle-resolved transport calculation \citep{Hwang2020AngleIndependent} is the
natural extension of the present framework. The thick, low-contrast pigment
films of \cite{Li2025Nexus}, whose angular stability is documented
photographically, lie in the strongly multiple-scattering limit that such a
calculation would address.

The sweeps above stay close to backscattering, since the observer and the
sun lie on the same side of the module normal or close to it. A façade is
often seen otherwise: the sun stands high above it and the observer looks up
at it from the street, on the opposite side of the normal, so the light
returned to the observer is scattered through smaller angles. Fig.~S10 follows
the four reference recipes along this family of geometries, with the sun and
the observer at the same polar angle $t$ on opposite sides of the normal, so
that the scattering angle inside EVA is $180^{\circ} - 2t_{\mathrm{in}}$. The
colors move away from their own normal-view colors by more than
$\Delta E_{00} = 5$ once $t$ reaches 14 to
24 degrees, and at $t = 50^{\circ}$ to
$60^{\circ}$, typical of a façade lit by a high sun and seen from below, they
have moved by $15$ to
$59$ units: the green renders as a
violet-blue, the gold as a green to cyan and the terracotta as a yellow-green,
while the blue darkens. The direction of these shifts follows from the
structure factor: the momentum transfer of Eq.~\eqref{eq:qtheta} scales with
$\sin(\theta_s/2)$, so the wavelength at which the correlated peak returns
light falls by the same factor as the scattering angle departs from
$180^{\circ}$, about $18\%$ at $\theta_s = 110^{\circ}$, and every color moves
towards the blue. Colors specified at normal view are therefore the colors of
a façade seen with the sun behind the observer; for the geometry a façade
mostly presents, the design target has to be set at the scattering angles the
building actually offers, a constraint that follows from the structure factor
itself rather than from the details of the present model.

\subsection{The daily cycle}\label{sec:daily}

\begin{figure*}[t]
  \centering
  \includegraphics[width=0.9\textwidth]{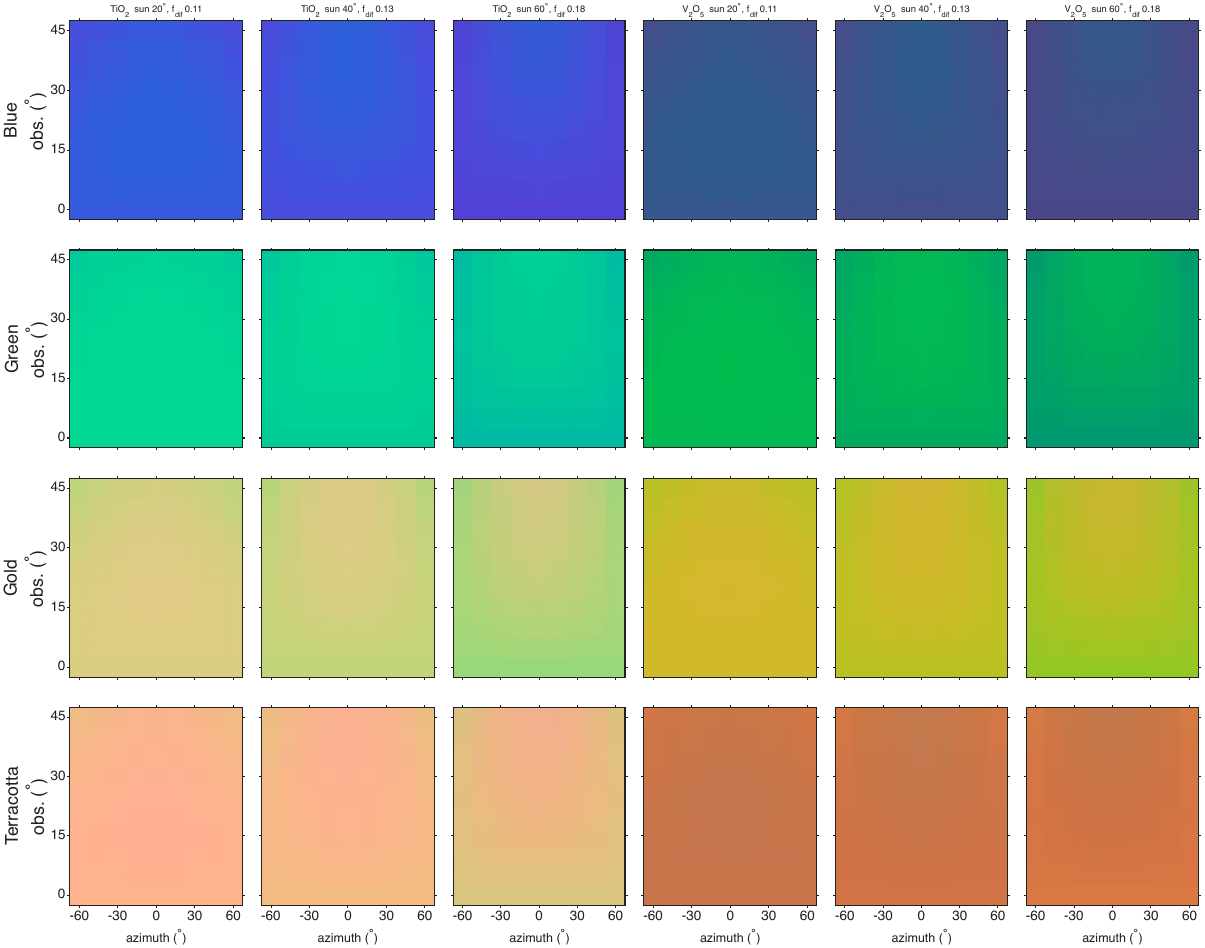}
  \caption{Rendered color over the solar cycle, with the diffuse fraction of
  a clear-sky model for the atmosphere of ASTM G173,
  $f_{\mathrm{dif}} = 0.111$, $0.129$ and
  $0.179$ at solar polar angles of $20^{\circ}$, $40^{\circ}$ and
  $60^{\circ}$ from the module normal. Each block sweeps observation polar
  angle and azimuth (in air). Columns one to three are TiO$_2$, columns four to
  six V$_2$O$_5$. Drift accumulated over all three solar positions ranges from
  $5.0$ for the V$_2$O$_5$ terracotta to
  $17.6$ for the V$_2$O$_5$ gold among the reference
  recipes.}
  \label{fig:vmix}
\end{figure*}

The preceding sections varied the illumination angle and the diffuse fraction
independently. In reality they are coupled: as the sun descends the beam
traverses a longer air mass, is attenuated more strongly, and the sky
contributes a growing share of the total irradiance.
Figure~\ref{fig:vmix} follows the recipes through this coupled variation, at
solar polar angles of $20^{\circ}$, $40^{\circ}$ and $60^{\circ}$ from the
module normal, with the diffuse fraction set by the clear-sky model of
Section~\ref{sec:stack}, giving $f_{\mathrm{dif}} = 0.111$,
$0.129$ and $0.179$ respectively,
and sweeping the observer over both polar angle and azimuth at each solar
position. The diffuse fractions are those of a horizontal surface and serve
here as a parametric guide; a vertical façade sees part of the ground and a
larger share of diffuse light.

Within a single solar position the drifts are comparable to those of
Section~\ref{sec:angular}: at a solar polar angle of $20^{\circ}$ the
reference recipes drift by $7.2$,
$7.3$, $9.6$ and
$3.0$ across the observer hemisphere. As the sun
moves further from the normal every recipe drifts more, to
$9.4$, $12.7$,
$14.4$ and $5.0$ at
$60^{\circ}$: the refracted beam moves further from the observer directions,
so the same sweep samples scattering angles further from backscattering, while
the diffuse share grows only from $0.11$ to
$0.18$, too little to pull the color towards the
hemispherical average.

Accumulating over all three solar positions, that is, taking the largest color
difference between any two viewing configurations at any of them, gives the
figure that matters for a façade seen at all hours. The reference recipes
accumulate $11.4$ (blue), $16.1$
(green), $17.6$ (gold) and
$5.0$ (terracotta). The terracotta, the tone most
frequently specified for colored façades together with green
\citep{ThreeColorFacade2026}, is thus also the most stable, for the reason
given above, whereas the green and the gold change by about as much as two
unrelated colors over the day. The closest TiO$_2$ counterparts, which do not
reach the green, gold and terracotta targets, accumulate
$14.0$, $18.0$ and
$24.8$, comparable for green and gold and far more
for terracotta, so the absorbing oxide that reaches these targets does not pay
for it in stability.

The design conclusion is that angular behavior must be checked alongside
color and set by the geometry the façade presents. Under a fixed sun and for
an observer moving near the normal, the reference colors hold to within about
a distinct-color step; over the day they drift by $5.0$
to $17.6$, terracotta least; and for a façade lit
from above and viewed from below they move towards the blue by far more than
that. The levers available are, in decreasing order of effect, the viewing
geometry, which the designer can shape through the orientation and shading of
the façade, the choice of a recipe whose color is set by an absorption edge
rather than a resonance, as for terracotta, and the coating thickness.

\section{Conclusion}\label{sec:conclusion}

We have developed a numerical framework for amorphous photonic-glass coatings
on photovoltaic modules and used it to ask which façade colors a metal-oxide
packing can produce, with which material, and how well those colors survive
the illumination and viewing conditions a building actually experiences.
Coupling Mie scattering, the Percus--Yevick structure factor, the exact Mie
phase function and two-flux multilayer transfer with absorption over a
photovoltaic substrate, with refraction at the front glass, colors evaluated
under AM1.5G and color differences in CIEDE2000, the model carries no fitted
optical parameter: the layer optical thickness follows from the packing
geometry, so every reflectance reported here is a consequence of particle
size, packing fraction, layer number and material alone.

The restriction to metal oxides is not a matter of convenience. Transparency
across the visible caps the refractive index near $2.35$ through the Moss
relation, and the materials that exceed that bound absorb within the band
they are meant to color. Metal oxides sit at the ceiling and are the class
from which monodisperse nanospheres can practically be grown, so the index
contrast available against a polymer encapsulant is modest and fixed at
roughly $1.7$ for TiO$_2$ and $1.3$ for the lower-index oxides.

Four findings follow. First, four target colors that a large crowd-sourced
survey names unambiguously blue, green, gold and terracotta are all reachable
in a laminated three-layer coating, to within $\Delta E_{00}$ of
0.8 to 2.5, and a saturated red is not.
Second, they are not reachable from one oxide. The transparent TiO$_2$ renders
the blue and only pale versions of the warm tones; the green, gold and
terracotta require the blue-band absorption of V$_2$O$_5$, which deepens them
at the cost of light that the cell would otherwise convert; the low-contrast
oxides reach blue only through a second-order resonance and turn grey at the
warm end. At the index ceiling, absorption is what extends a scattering
coating to the warm half of the palette. Third, the medium governs the result
and the substrate does not: the reference recipes shift by $\Delta E_{00}$ of
17.6 to 45.1 between the unlaminated and laminated
states, because the encapsulant relocates the resonance, whereas exchanging a
PERC cell for a TopCon one moves them by less than half a unit. Fourth, angle
independence is not conferred by disorder alone. Refraction at the front glass
compresses the angles inside the stack, and under a fixed sun the colors hold
for an observer moving near the normal, but over the day the reference recipes
drift by $\Delta E_{00}$ of 5.0 to
17.6, and in the geometry of a façade lit from
above and seen from below every color moves towards the blue by far more,
because the correlated peak shifts with $\sin(\theta_s/2)$. The terracotta,
whose color is set by an absorption edge, is the most stable of the four;
thickness redistributes stability between tones but does not remove the drift.

Taken together these convert a qualitative expectation into a design
procedure. A coating should be specified for the laminated stack from the
outset, which is also where glass-side deposition delivers it; the oxide
should be chosen for the target color, a transparent high-index oxide for the
cool tones and a blue-absorbing one for the warm tones and green, with the
photocurrent cost of the absorption weighed against the color; the layer
number should be set where all targets keep their names and tolerate
process variation, here $N = 3$; the color should be specified at the
scattering angles that the façade presents rather than at normal view; and the
cell technology beneath can be ignored.

Four directions extend this work. The two-flux treatment keeps the model free
of fitted parameters and inexpensive enough to map the full design space, but
it has not been benchmarked against an angle-resolved transport calculation,
which would add the redistribution of direction within the stack and to which
the thickness and angular results are the most sensitive. The optical
constants are taken from published determinations on films and single
crystals rather than measured on the specific particles a manufacturer would
use, and the sensitivity analysis in the Supplementary Material shows what
this means in practice: a five percent change in the refractive index of the
oxide displaces the reference colors by $\Delta E_{00}$ of
2.6 to
6.0 at a fixed diameter, of which
retuning the diameter recovers all but
1.1 to
5.9; for V$_2$O$_5$ the absorption
is taken from published fits and figures of a single film preparation, and
indices reported for sputtered films are higher by up to a quarter of a unit
in the red. The diameters quoted here are therefore design starting points,
and the qualitative conclusions are the robust result. The lamination shift could be removed by locking the oxide packing in a
polymer matrix and grinding it into grains, as already done for silica
\citep{Li2025Nexus}, which would decouple the color from the encapsulant. Finally, the
predictions, including the blue shift in the façade geometry, provide concrete
targets for experimental tests on sprayed coatings.

\section*{Data and code availability}
 
The model code, the optical constants and every other input, the data
behind each figure and the MATLAB scripts that draw them are openly
available at Zenodo, \url{https://doi.org/10.5281/zenodo.23196090}, the code
under the MIT license and the data under CC~BY~4.0.

\section*{Acknowledgments}

This work was supported by the Scientific and Technological Research Council of Turkey (TÜBİTAK) under grant nr.\ 123F372.

\printcredits

\section*{Declaration of competing interest}
The authors declare that they have no known competing financial interests or personal relationships that could have appeared to influence the work reported in this paper.
   
\section*{Declaration of generative AI and AI-assisted technologies in the writing process}
During the preparation of this work, the authors used Claude (Anthropic) solely to improve the language, readability, and flow of the manuscript text. The tool was not used to generate scientific content, data, results, figures, or interpretations. After using this tool, the authors reviewed and edited the content as needed and take full responsibility for the content of the publication.

\bibliographystyle{elsarticle-num}

\bibliography{mainrefs}



\end{document}